\documentclass[lettersize,journal]{IEEEtran}
\IEEEoverridecommandlockouts
\usepackage[compact]{titlesec} 
\usepackage{amsmath,amsfonts}
\usepackage{booktabs}
\usepackage{fancyvrb}
\usepackage{tabularx}
\usepackage{multirow}
\usepackage{color}
\usepackage{float}
\usepackage{cite}
\usepackage{enumitem}  
\usepackage[utf8]{inputenc}
\usepackage{lmodern}
\usepackage{tabularx}
\usepackage{csquotes}
\usepackage[justification=centering]{caption}
\usepackage{blindtext}
\usepackage{wrapfig}
\usepackage{graphicx}
\DeclareGraphicsExtensions{.eps,.pdf,.gif,.png,.jpg}
\usepackage{amssymb}
\usepackage{amsthm}
\usepackage{array}
\usepackage{algorithm,algorithmic}
\usepackage{booktabs}
\usepackage{graphicx}
\usepackage{epstopdf}\usepackage[utf8]{inputenc}
\usepackage[hyphens,spaces,obeyspaces]{url}
\usepackage[english]{babel}
\usepackage{verbatim} 
\usepackage{lineno,nohyperref}
\usepackage{epstopdf} 
\modulolinenumbers[5]

\RequirePackage{filecontents}
\usepackage{url} 
\usepackage[T1]{fontenc}
\usepackage[utf8]{inputenc}
\usepackage{microtype} 

\newcommand*{\affaddr}[1]{#1} 
\newcommand*{\affmark}[1][*]{\textsuperscript{#1}}
\newcommand{\vect}[1]{\boldsymbol{#1}}
\newcommand*{\email}[1]{#1}
\PassOptionsToPackage{hyphens}{url}
\newtheoremstyle{mystyle}
{}
{}
{\itshape}
{}
{\bfseries}
{.}
{ }
{\thmname{#1}\thmnumber{ #2}\thmnote{ (#3)}}

\theoremstyle{mystyle}
\newtheorem{remark}{Remark}

\newcounter{subassumption}[asu]

\makeatletter
\renewcommand{\p@subassumption}{\theasu}
\makeatother
\usepackage{amsthm}
\usepackage{xpatch,amsthm}
\makeatletter
\xpatchcmd{\@thm}{\fontseries\mddefault\upshape}{}{}{} 
\makeatother

\makeatother
\usepackage{cite}
\usepackage{amsmath,amssymb,amsfonts}
\usepackage{algorithmic}
\usepackage{graphicx}
\usepackage{textcomp}
\usepackage{xcolor}
\def\BibTeX{{\rm B\kern-.05em{\sc i\kern-.025em b}\kern-.08em
		T\kern-.1667em\lower.7ex\hbox{E}\kern-.125emX}}

\begin{document}	
	\title{Empowering Rural Areas with Multi-radio Microwave Backhaul  Supported by Digital Twin for 5G IAB-based FWA \\
	\thanks{This work was supported by NSERC (under project ALLRP 566589-21) and  InnovÉÉ (INNOV-R program) through the partnership with Ericsson and ECCC.
		We thank the Ericsson's Montréal
		GAIA team for their constructive and helpful comments, which have
		significantly improved the quality and clarity of this manuscript (corresponding author: Anselme Ndikumana).}}
		\author{%
	Anselme Ndikumana\affmark[1], Kim Khoa Nguyen \affmark[1], Adel Larabi\affmark[2], and Mohamed Cheriet\affmark[1]\\
	\affaddr{\affmark[1]Synchromedia Lab, École de Technologie Supérieure, Université du Québec, QC, Canada\\  \email{\{anselme.ndikumana, kim-khoa.nguyen; Mohamed.Cheriet\}}@etsmtl.ca}\\
	\affaddr{\affmark[2] GAIA  Montreal, Ericsson Canada}\\
	\email{\{adel.larabi\}@ericsson.com}\\
	}
	\maketitle
	
	\begin{abstract} 
For digital inclusion, high-capacity Internet access should be provided to rural areas to support a range of services and applications. Due to the high operating costs of fiber-optic deployment, Fixed Wireless Access (FWA) is becoming a more attractive internet solution for rural areas. However, 5G FWA is a one-hop solution with limited coverage. A multi-hop solution is needed for wider rural coverage. This work considers a unified solution combining long-haul microwave, 5G Integrated Access and Backhaul (IAB), and FWA to provide a multi-hop network for extended coverage and high network capacity in rural areas. A key challenge for such a network is that energy consumption increases with the number of hops, a problem that has been overlooked in the existing literature. To address this, we propose energy-efficiency microwave backhaul for IAB-based FWA as the Physical Twin (PT). We develop an energy-efficient strategy to optimize radio start-up, serving, sleeping, and wake-up states for microwave backhaul connecting 5G IAB-based FWA serving rural areas. By operating the network at reduced capacity during low utilization, we aim to minimize energy consumption. Then, we present a Digital Twin (DT) of PT to improve its performance. We solve the formulated optimization problem using deep Q-learning in DT and the optimization solver in PT. The simulation results show that our approach satisfies the data rate requirements while reducing energy consumption.
	\end{abstract}
	
\section{Introduction}
\subsection{Background and Motivations}
\label{subsec:motivation} 
Access to high-speed internet, particularly 5G networks, remains uneven worldwide. Many regions, especially rural areas, are either unconnected or rely on slower network technologies such as 2G and 3G. According to the International Telecommunication Union (ITU), 2.2 billion people remain offline in 2025 \cite{ITU}. Expanding high-speed internet access in rural areas is essential for fostering economic growth, improving education, and enhancing the quality of life. While fiber optic deployment offers high reliability and bandwidth, it is often cost-prohibitive in low-density regions with uncertain returns on investment. As a cost-effective alternative, 5G Fixed Wireless Access (FWA) enables connectivity by equipping homes with rooftop Customer Premises Equipment (CPE) that wirelessly links to fixed cellular base stations \cite{Maravedis}. The number of FWA connections is projected to increase from $160$ million by the end of $2024$ to $350$ million by $2030$, representing 19\% of all fixed broadband connections \cite{ericsson2024Mobility}.

\begin{figure}[t]
	\centering
	\includegraphics[width=1.0\columnwidth]{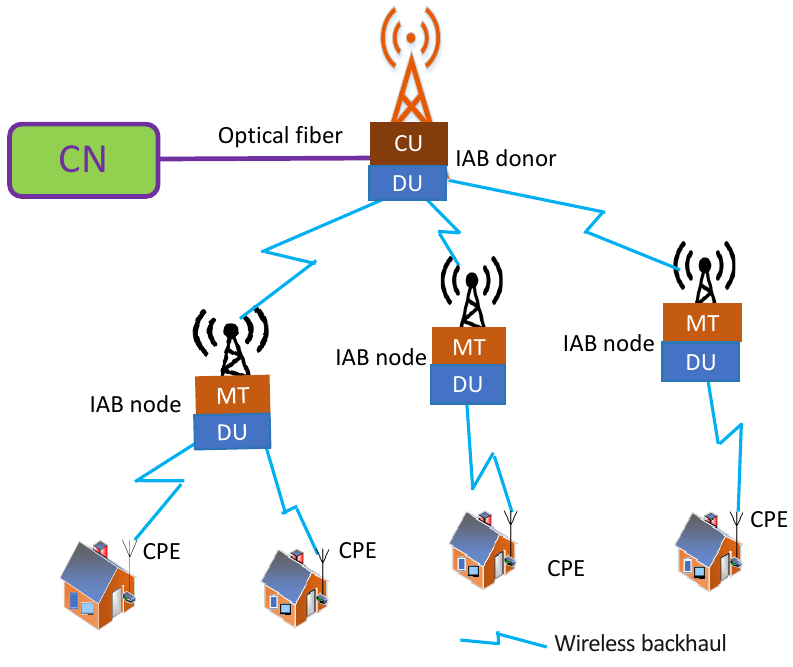}
	\caption{Illustration example of 5G IAB-based FWA.}
	\label{fig:top_IAB_FWA}
\end{figure}

Mid-band and mmWave frequencies help to enhance the capacity of FWA networks \cite{chaudhuri2021extended}. However, 5G FWA is a single-hop solution, which limits its coverage range. To extend coverage beyond a single hop, additional technologies are needed to enable multi-hop FWA. In \cite{hashemi2017integrated}, the authors proposed a unified solution that considers Integrated Access and Backhaul (IAB) in FWA to cover larger areas. An IAB network \cite{3GPP38874} consists of an IAB donor and multiple IAB nodes. The IAB donor is a base station connected to the Core Network (CN) via a fiber-optic link, while the IAB nodes are additional base stations that connect to the donor via wireless backhaul links. As shown in Fig. \ref{fig:top_IAB_FWA}, in the IAB-based FWA deployment, CPEs can connect to either the IAB donor or the IAB nodes using wireless links. Each IAB node comprises a Distributed Unit (DU) and a Mobile Termination (MT) unit. The DU serves both CPEs and downstream MT. The MT allows the node to operate as a relay node when connecting to its parent DU. The IAB donor includes both a DU and a Centralized Unit (CU).

Deploying fiber-optic backhaul for an IAB donor is often not cost-effective, particularly in rural areas with low population density and uncertain returns on investment. To address this challenge, we consider using microwave backhaul as a practical and efficient alternative for connecting an IAB donor to the core network for internet access. Long-haul microwave radios operating in frequency ranges such as $11$ $GHz$, $71-86$ $GHz$, and $191.7-194.8$ $THz$ have been tested in rural areas, particularly agricultural zones with limited broadband access \cite{zu2023arahaul}. Despite its potential, the integration of long-haul microwave, FWA,  and IAB as a unified network framework remains relatively unexplored in the literature. By integrating long-haul microwave, IAB, FWA, and high-frequency bands as a unified network, it becomes possible to extend coverage across vast rural areas while maintaining high data rates. 

We can use a Digital Twin (DT) to model a microwave backhaul connecting an IAB-based FWA network to the internet, where the DT continuously reflects the physical microwave's processes, dynamics, and states. In other words, the DT serves as a virtual representation of the Physical Twin (PT), while the actual microwave constitutes the PT. DT \cite{tao2022digital} can play a transformative role in enhancing microwave backhaul service in rural areas. Instead of sending technicians to rural areas for every microwave backhaul malfunction, many network issues, such as remote antenna alignment or configuration changes, can be diagnosed and addressed remotely using the DT, significantly reducing operational costs and response times while improving microwave backhaul service reliability.

\subsection{Challenges for Deploying Long-haul Microwave and 5G IAB-based FWA}
\label{subsec:challenges} 
Long-haul microwave and multi-hop IAB-based FWA networks present several critical challenges when deployed to serve rural areas:
\begin{itemize}
	\item Combining long-haul microwave with IAB-based FWA offers a promising solution for extending broadband coverage in rural areas. However, energy consumption tends to increase with the number of hops, and current research \cite{zhang2021resource, yu2023coordinated, zu2023arahaul, begishev2021performance} has not yet addressed this issue.
	\item In multi-radio{\tiny } microwave backhaul, determining which radios to place in deep sleep mode during periods of low traffic, and when to reactivate them as demand increases, is a complex task. Efficient radio management is essential to achieving energy savings without degrading network performance \cite{frithiofson2022energy}.
	\item Energy-saving techniques, such as deep sleep, can adversely affect network performance during unexpected traffic surges. Delays in waking up radios may lead to service degradation or increased latency \cite{frithiofson2022energy}.
	\item Rural areas typically have lower population density compared to urban and suburban environments. As a result, radio resources allocated to CPEs and MTs often remain underutilized, leading to inefficient use of resources \cite{ndikumana2024renewable}.
	\item  Providing on-site technical support in rural areas is often logistically difficult and cost-prohibitive. Therefore, network automation using DT should be considered to enable remote network monitoring, diagnostics, and fault management. Despite its potential, the application of DT technology in rural connectivity scenarios is still largely underexplored in the existing literature.
\end{itemize}

\subsection{Contributions}
\label{subsec:contribution} 

In this paper, we propose an energy-efficient microwave backhaul framework, supported by DT, to connect an IAB-based FWA network serving rural areas to address the key challenges discussed above. The main contributions of this paper are summarized as follows:
\begin{itemize}
	\item We propose a multi-radio microwave backhaul system where each microwave node and IAB donor is equipped with multiple microwave radios. These radios, considered part of the Physical Twin (PT), can dynamically enter low-power states to reduce energy consumption during periods of low network utilization. Unlike existing literature, which typically models radios in only three states, off, on, and deep sleep, and ignores the latency associated with transitioning between these states, we introduce a more realistic five-state model: completely off, startup, serving, deep sleep, and wake-up. We then define state and action spaces to manage state transitions while minimizing energy consumption.
	\item We develop a DT model for the microwave backhaul that receives network metrics from the PT. The DT uses deep Q-learning (DQL) \cite{ernst2024introduction} to optimize state transitions of the microwave radios, aiming to minimize energy consumption when the network is underutilized. The optimized transition matrix (i.e., action-state mapping) is then sent to the PT to guide its operations.
	\item Upon receiving feedback from the DT, at the PT, the PT uses an optimized transition matrix to minimize energy consumption based on Age of Processing (AoP), while ensuring the data rate requirements of the IAB-based FWA serving rural areas are met. We solve this problem using optimization solver.
\end{itemize}

In this work, we adopt the AoP metric \cite{li2021age} to evaluate the timeliness of the PT's sending of network metrics to the DT. The DT then optimizes radio states, sends the optimized states as feedback to the PT, and the PT subsequently uses this feedback to improve its operations. AoP is an extension of the Age of Information (AoI). This metric represents the time elapsed between the generation of a status at the source node and its most recent update at the destination node \cite{yates2021age}. However, in practical systems, such as microwave backhaul, useful status information can be obtained only after processing the collected data from the PT, e.g., computation or inference. Therefore, AoP incorporates this additional computation delay into the AoI framework, offering a more comprehensive view of end-to-end freshness. It has been applied in various real-time applications, including data sampling, offloading, and processing in Internet of Things (IoT) systems \cite{li2021age}, as well as data offloading for vehicular networks \cite{ndikumana2022age, ndikumana2023age}.

The remainder of the paper is organized as follows: Section \ref{sec:literaturereview} reviews related work, and Section \ref{sec:SystemModel} presents the system model. In Section \ref{sec:problem_formulation}, we formulate the optimization problem, and Section 
\ref{SolutionApproach} outlines the proposed solution. Section \ref{sec:PerformanceEvaluation} includes the performance evaluation. Finally, we conclude the paper in Section \ref{sec:Conclusion}.

\section{Literature Review}
\label{sec:literaturereview} 
We categorize the existing related works into three main groups: (i) microwave backhaul and radio link bonding,  (ii)  FWA and resource allocation supported by DT, and (iii)  IAB-based FWA and resource allocation.

\emph{ Microwave Backhaul and Radio-link Bonding:}
Microwave is widely used across a range of frequency bands and is expected to remain a vital transport technology for 5G networks, as noted in \cite{sellin2020enhancing}. In \cite{zu2023arahaul}, long-haul microwave radios operating in the 11 GHz, 71–86 GHz, 191.7–194.8 GHz spectrum were evaluated for use in a rural agricultural region lacking broadband connectivity. A key technique for increasing capacity in such a network is radio-link bonding, which aggregates data across multiple frequency carriers.
The authors in \cite{wang2022triple} explored triple-band scheduling involving the 28 GHz band, the E-band (71–76 GHz paired with 81–86 GHz), and the Terahertz (THz) band. The integration of these high-frequency bands is better suited to high-density urban environments, given their limited coverage and high path loss.
In \cite{colzani2022long}, the potential of high-power amplifier modules was examined for enhancing long-reach E-band systems in a radio-link bonding setup, improving backhaul performance for long distances. The authors in \cite{frithiofson2022energy} pointed out that microwave backhaul links are often underutilized, with usage rarely exceeding 50\%. This underutilization presents an opportunity to reduce energy consumption by powering down some microwave radios during periods of low demand.
However, managing energy efficiency in the microwave network with multiple radios and multi-band configurations presents significant challenges. Specifically, it is difficult to determine optimal times and conditions for putting specific radios or carriers into deep sleep or awakening them, ensuring that traffic demands are met without network service disruption when capacity needs suddenly increase.

\emph{FWA and Resource Allocation Supported by DT:} 
In \cite{rahmawati2022assessing}, the authors proposed a method to determine the optimal number of base stations required for FWA capacity and coverage planning, using an urban residential area as a case study. Similarly, in \cite{lappalainen2022planning}, the authors introduced an approach to estimate the maximum number of houses that can be simultaneously connected to an FWA network while meeting minimum target bit rates, based on available network resources and cell radius. The work in \cite{de2022outdoor} focused on mmWave channel modeling for a 5G FWA network operating at 60 GHz. Furthermore, \cite{de2023mmwave} analyzed FWA performance across multiple frequency bands, specifically 28 GHz, 60 GHz, and 140 GHz, highlighting trade-offs between higher-frequency path loss and the benefits of larger bandwidths, which enable higher channel capacity. In \cite{castellanos2023evaluating}, the authors evaluated the deployment of 60 GHz FWA. This study collectively emphasizes the potential of high-frequency bands in delivering high-capacity FWA services. For rural connectivity, the authors in \cite{ndikumana2024digital, ndikumana2023digital} proposed using digital twins and closed-loop systems to manage radio resource allocation in FWA networks. Although promising, the proposed approaches are also limited to single-hop FWA networks.

\emph{ IAB-based FWA and Resource Allocation:}
In \cite{hashemi2017integrated}, the authors proposed an integrated approach combining IAB and FWA in the high-frequency 28 GHz band to deliver high-capacity connectivity to residential homes. In \cite{tafintsev2023airborne}, the use of unmanned aerial vehicles as IAB nodes was explored to enhance the flexibility and adaptability of IAB network topologies.
Given that high-frequency bands are prone to significant path loss, especially when obstacles are present, \cite{begishev2021performance} proposed a multi-band solution combining microwave and millimeter-wave in 5G new radio systems. Their approach mitigates outages by instantly rerouting traffic to sub-6 GHz links when mmWave connections become unavailable. Similarly, \cite{yao2022delay} introduced an architecture that integrates mmWave and sub-6 GHz bands to address mmWave blockages and intermittent connectivity. They focused on packet scheduling strategies that leverage both interfaces to maintain service reliability.
In rural environments, where the population density is low, radio resources allocated to CPEs and MTs are often underutilized. Addressing this, \cite{yu2023coordinated} highlighted the importance of radio resource allocation coordination in IAB networks, which can be either distributed (managed by individual IAB nodes) or centralized (managed by the IAB donor). They proposed a hybrid coordination method that combines both centralized and distributed strategies. However, their approach does not address the increasing energy consumption and latency that occur as the number of hops in the IAB network grows. In summary, while existing studies focus on the design of IAB-based FWA networks and radio resource allocation strategies, the challenge of minimizing energy consumption in multi-hop IAB-based FWA networks, particularly for rural deployments, remains largely unaddressed in the literature.

Based on the related works reviewed above, the integration of long-haul microwave supported by DT with multi-hop IAB-based FWA for rural connectivity has not been thoroughly explored in the existing literature. To the best of our knowledge, this is the first study to focus on minimizing energy consumption in such networks by leveraging multi-radio microwave backhaul supported by DT to connect IAB-based FWA serving rural areas.

\section{System Model}
\label{sec:SystemModel} 
	 \begin{figure}[t]
		\centering
\includegraphics[width=1.0\columnwidth]{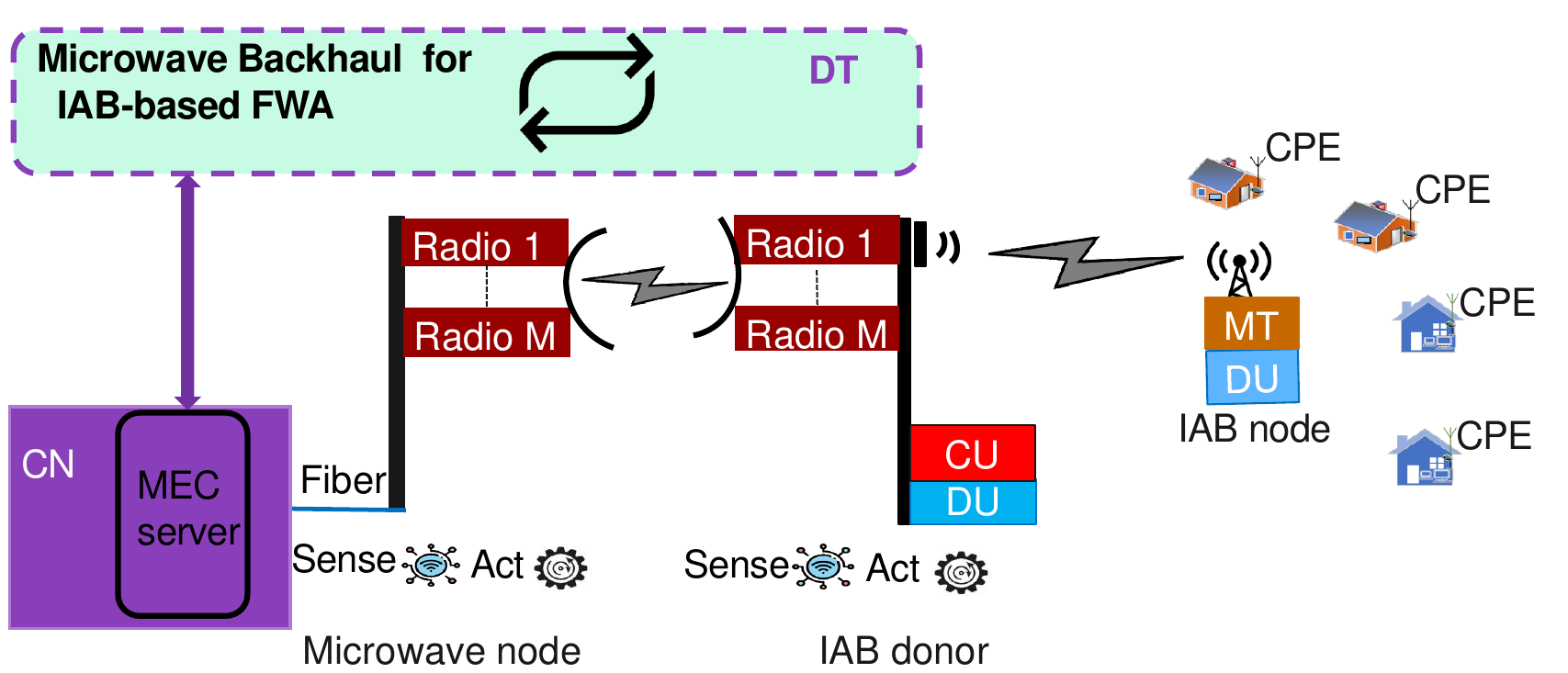}
		\caption{Illustration of the system model.}
		\label{SystemModelF}
	\end{figure}
\begin{table}[t]
	\caption{Summary of key notations.}
	\label{tab:table1}
	\begin{tabular}{ll}
		\toprule
		Notation & Definition\\
		\midrule
		$\mathcal{M}$ & Set of microwave radios  $|\mathcal{M}|= M$\\
		$\mathcal{J}$ & Set of microwave nodes  $|\mathcal{J}|= J$\\	
		$\mathcal{V}$ & Set of CPE and IAB-MT  $|\mathcal{V}|= V$\\
		$D$ & DL data rate requirement\\
		$K$ & Total number of states\\
		$\mathcal{S}^j$ & States of node $j$\\
        $\mathcal{A}^j$ & Actions of node $j$\\
         $P^j_{{m,k}}$ & Power consumption for radio $m$ in state\\
         & $k$ at node $j$ \\
		$P^{TX}_{m,k}$  & Transmission power  for radio $m$ in state $k$ \\
		$G^{RX}_{m,6} $ & Channel gain\\
		$\Tilde{D}_{rds}$&  Deep sleep threshold\\
		$\Tilde{D}_{w}$ &  Radio wake-up threshold:\\
		$E_m^i$ & Time between sending update $i$ to DT and \\
         &receiving feedback\\
         $\Phi(\mathcal{S}^j, \mathcal{A}^j)$& State transition matrix \\
		$\tilde{B}_{m}$ & The average AoP\\
		$U_m$ & Time of freshest status update at the PT $m$ \\
		$\vect{x}$ & Vector of  state selection variables\\
		$\vect{y}$ & Vector of KPI/SLA  satisfaction variables at PT \\
		$\vect{z}$ & Vector of KPI/SLA  satisfaction variables at DT \\
		$H_m^{i2}$ & Area of triangle \\
		$H_m^{i1}$ & Area of parallelogram\\
		$\Delta_\tau$ & Desynchronization time between DT and PT\\
		$L_{mec}$ & Energy consumption of DT \\
		$R(\mathcal{S}^j, \mathcal{A}^j)$& Reward function in DT\\
		$D^j_m$& Achievable data rate for radio $m$ of node $j$ \\ 
		\bottomrule
	\end{tabular}
\end{table}
\begin{figure}[t]
\centering	\includegraphics[width=0.8\columnwidth]{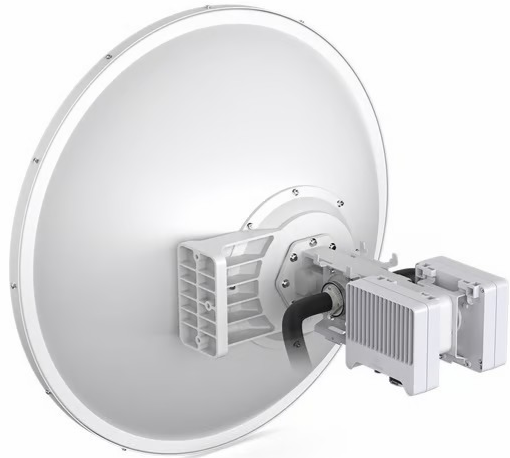}
	\caption{Illustration example of an antenna with two microwave radios \cite{ericssonant}.}
	\label{fig:microwavetwol}
\end{figure}
\begin{figure}[t]
	\centering	\includegraphics[width=1.0\columnwidth]{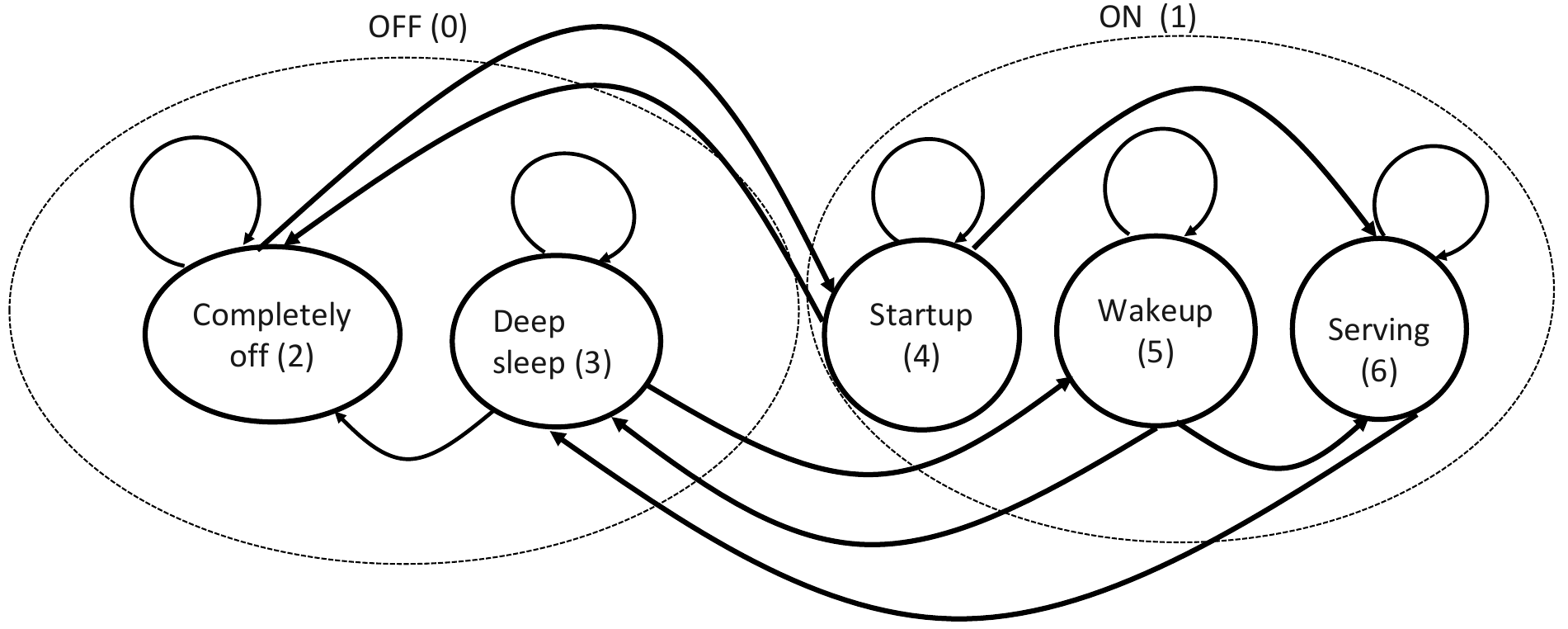}
	\caption{Illustration of microwave radio states.}
	\label{fig:mdp_Model}
\end{figure}

We illustrate our system model in Fig.~\ref{SystemModelF}, while the key notations used in this paper are summarized in Table~\ref{tab:table1}.

Our system model considers the microwave backhaul between a microwave node and the IAB donor. The microwave node is connected to the Core Network (CN) via optical fiber. Each microwave node and IAB donor is equipped with more radios. Fig. \ref{fig:microwavetwol} shows an example of a multi-band booster antenna with two radios, which can be attached to the microwave node and the IAB donor. Let $\mathcal{M}$ denotes a set of microwave radios and $\mathcal{J}$ a set of microwave nodes and IAB donor, where $M^j$ represents the number of radios associated with microwave node or IAB donor $j \in \mathcal{J}$. We consider multi-band microwave radios that can operate in narrow (e.g., $6$ – $15$ $GHz$), wide (e.g., $18$ – $42$ $GHz$), or very wide (e.g., $71$ – $76$ $GHz$ and $81$ – $86$ $GHz$) frequency bands, depending on the data rate requirements of IAB-based FWA serving rural area.

In IAB-based FWA, we consider $\mathcal{V}$ to be the set of CPEs and IAB MTs that use microwave backhaul. We denote $d_v$ as the downlink (DL) data rate requirement for terminal $v$, i.e., CPE or IAB MT. Hereafter, the term terminal denotes either a CPE or an IAB-MT. Since we have multiple terminals, the DL data rate requirement can be denoted as:
\begin{equation}
	D =\sum_{v=1}^{V} d_v.
	\label{eq:data_rate_requiremnet}
\end{equation} 
The data rate $D$ passes through the microwave node and the IAB donor, each of which has multiple radios.

We model the microwave backhaul connecting the IAB-based FWA network to the CN as a physical twin (PT). A corresponding DT is deployed on a Multi-access Edge Computing (MEC) server \cite{ndikumana2019joint} in CN to represent the microwave backhaul. The DT continuously receives network metrics from the PT and optimizes the radios' state transitions to minimize energy consumption during periods of low network utilization. The resulting optimized transition matrix (i.e., action-state mapping) is then communicated back to the PT as feedback. Upon receiving feedback from the DT, the PT  uses it to minimize both the AoP and energy consumption while meeting the IAB-based FWA's data rate requirements.

In IAB-based FWA, the IAB donor is equipped with microwave radios for backhauling, as well as other radio(s) to serve IAB nodes and CPEs for home internet access. We use the term IAB station to refer to either an IAB donor or an IAB node. Each IAB station supports dual-band operation with mid-band and millimeter-wave (mmWave) interfaces. Here, we remind that in IAB-based FWA, the DU has a MAC scheduler that allocates resource blocks (RBs). Since each IAB station includes an O-DU, we use the terms DU and IAB station interchangeably unless otherwise specified. The proposed  DT and multi-state approach are applied exclusively to the microwave backhaul radios, while the access radios serving IAB nodes and CPEs are considered to operate continuously.  Here, we focus on modeling microwave backhaul for connecting IAB-based FWA, while the IAB-based FWA modeling, including self-backhauling, is discussed in our previous works in \cite{ndikumana2025energy, ndikumana2026energy}.

\subsection{Microwave Backhaul for Connecting IAB-based FWA}

As shown in Fig. \ref{fig:mdp_Model}, we consider two physical states for the microwave radio: OFF (denoted $0$) or ON (denoted $1$). Also,  the microwave node has a processor unit equipped with the controller. At the controller, we consider five states: completely off (denoted $2$), deep sleep (denoted $3$), startup (denoted $4$), wakeup (denoted $5$), and serving state (denoted $6$). When combining the physical states and controller states, we use $K$ to denote the total number of states.
The state $\{ 0\{2, 3\}$  and $\{ 1\{4, 5, 6 \}\} \times \{1,2, \dots, M^j\}$ can be  represented as:
\begin{equation}
	\mathcal{S}^j=\{\mathcal{S}^j_{m, k}\}| m=1, 2, \dots, M^j, k=0, 1, 2, \dots, K.
\end{equation}
In the OFF(0) state, we distinguish between deep sleep and being completely off. Completely off refers to a unit that is either new or unused in a serving state for a long period, such as consecutive days. In a deep sleep state, we consider time on a small scale rather than on a daily scale. In the ON(1) state, the controller considers startup ($4$), wake-up ($5$), and serving ($6$) states.

We denote  $P^j_{{m,k}}(t^j_m)$ as the power consumption of microwave radio $m$ of node $j$ at the state $k$ at time $t^j_m$. Also, we define $\vect{x}_j=\{x^j_{m, k} \}$ as a vector of decision variables that indicate the state microwave radio belongs to. The decision variable $x^j_{m, k}$ is defined as:
\begin{equation}
	\setlength{\jot}{10pt}
	x^j_{m, k} =
	\begin{cases}
		1\;\text{if microwave radio $m$ of microwave node $j$}\\
		\;\;\;\text{is in the state $k$,} \\
		0, \; \text{otherwise.}
	\end{cases}
\end{equation}

To satisfy the data rate requirement $D$  that passes through the microwave backhaul, let us consider $m$ and $n$ as two microwave radios. Radio $n$ is located at the microwave node, and radio $m$ is located at the IAB donor.   Then, we  express the achievable SNR $\delta_{m}$ at microwave radio $m$ as follows:
\begin{equation}
	\label{eq:AchSNR}
	\delta_{m} = 
	\frac{|G^{RX}_{m,6} |^2  P^{TX}_{n,6} }{\sigma_{mn}^2},
\end{equation}
where $\sigma_{mn}^2$ is the noise power,  $P^{TX}_{n,6}$ is the transmission power in serving state, and $G^{RX}_{m,6}$ is the channel gain for microwave in ON and serving states. Based on the achievable SNR, the maximum achievable DL data rate for microwave radio $m$ of IAB donor $j$ is given by:
\begin{equation}
	\label{eq:data_rate}
	\begin{aligned}
		D^j_m=x^j_{m, 6}\omega_m^{b}\log_2\left(1 + \delta_{m} \right), 
	\end{aligned}
\end{equation}
where $\omega_m^{b}$ is the channel bandwidth. Considering all microwave radios $M^j$ used at IAB donor $j$, the data rate should satisfy:
\begin{equation}
	\label{eq:datarequirement}
	\sum_{m=1}^{M^j} 
	 D^j_m \geq D,
\end{equation}

At the time $t^j_m$, when the existing microwave radios in the serving state can not satisfy (Eq. \ref{eq:datarequirement}), the controller considers switching the existing microwave radio(s) from the completely off state to the startup state. 
Therefore, we define an action $a^{1,j}_{m,4}$ of switching from completely off to the startup state. Otherwise, action $\tilde{a}^{0,j}_{m,2}$ is taken, where the microwave radio remains in a completely off state. By taking action $a^{1,j}_{m,4}$, the controller sends a command to the microwave radio to switch physically from the OFF state to the ON state. This action is denoted $a^{1,j}_{m}$. When the microwave radio starts, it stays in the startup state for
\begin{equation} 
	\sum_{t^j_m=1}^{\tau}x^j_{m, 4}t^j_m \leq t^j_{m,4},
\end{equation} 
where $t^j_{m,4}$ is the time required for the microwave radio to complete the startup process. In other words, microwave radio takes action  $\tilde{a}^{1,j}_{m,4}$ to stay in the startup state until $t^j_{m,4}$ expires.  If the startup process is not completed within startup time $t^j_{m,4}$,  the microwave radio returns to the completely off state by taking action $\tilde{a*}^{0,j}_{m,2}$ and tries to start again. Otherwise, the controller considers switching the microwave radio from the startup state to the serving state by taking action $a^{1,j}_{m,6}$.
\begin{figure*}[t]
	\centering
	\includegraphics[width=2.0\columnwidth]{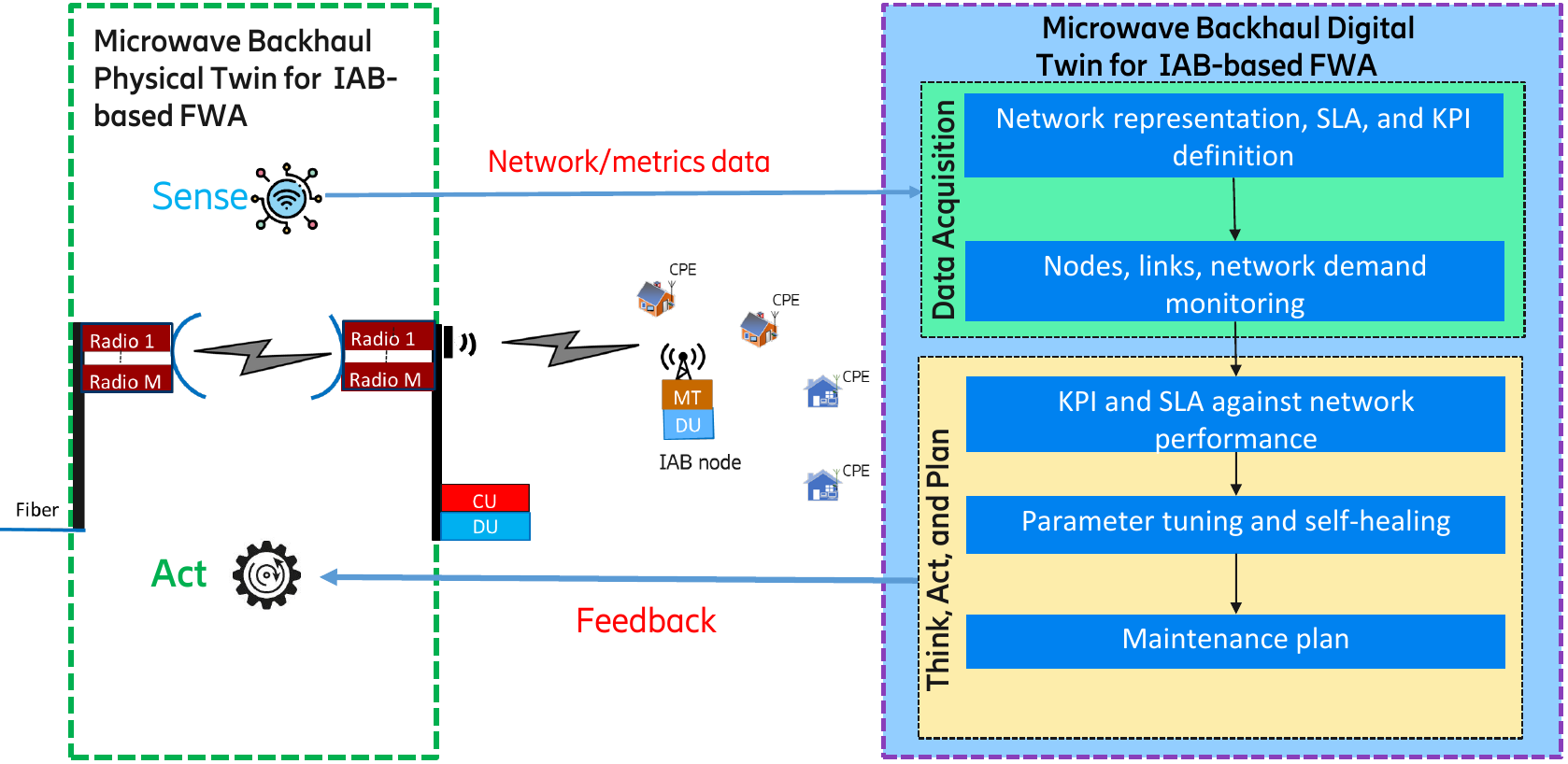}
	\caption{Interaction between PT and DT of microwave backhaul.}
	\label{SystemModel2}
\end{figure*}

When network traffic falls below a certain threshold, certain serving microwave radios can enter a deep sleep state to reduce energy consumption. Therefore, we define the following data traffic $\Tilde{D}_{rds}$ as the deep sleep threshold at the microwave node or at the IAB donor:
\begin{equation}
	\Psi_{rds} =\max \{(\sum_{m=1}^{M^j} D^j_m (t^j_m) -D (t^j_m)), \Tilde{D}_{rds} \}.
\end{equation}
When $\Psi_{rds} = \Tilde{D}_{rds}$, we consider switching from the serving state to the deep-sleep state for microwave radio in the ON state. We denote this action as $a^{0,j}_{m,3}$. By taking action $a^{0,j}_{m,3}$, the controller sends a command to the microwave radio to switch physically to the OFF state. This action is denoted $a^{0,j}_m$.   Otherwise, microwave radio remains in the serving state by taking action $\tilde{a}^{1,j}_{m,6}$.

When backhaul traffic increases again, a microwave radio in a deep sleep state can wake up. Therefore, we define the following data traffic $\Tilde{D}_{w}$ as the radio wakeup threshold at the microwave node or at the IAB donor:
\begin{equation}
	\Psi_{w} =\max \{(\sum_{m=1}^{M^j} D^j_m (t^j_m) -D (t^j_m)), \Tilde{D}_{w} \}.
\end{equation}
When $\Psi_{w} = \Tilde{D}_{w}$, we consider switching from the deep sleep state to the wakeup state. We denote this action as $a^{1,j}_{m,5}$. Otherwise, microwave radio remains in a deep sleep state by taking action $\tilde{a}^{0,j}_{m,3}$. We consider $ t^j_{m,5}$ as the time required so that the microwave radio completes the wakeup process, such that:
\begin{equation}
	\sum_{t=1}^{\tau}x^j_{m, 5}t^j_m \leq t^j_{m,5}.
\end{equation}
When a microwave radio is in a startup state and cannot complete the wakeup process within the startup time $t^j_{m,5}$, it fails, returns to the deep sleep state, and tries again. This action is denoted $\tilde{a*}^{0,j}_{m,3}$. Otherwise, the microwave radio switches from the wakeup state to the serving state. This action is denoted $a^{1,j}_{m,7}$.

When the microwave radio is in deep sleep for a long period $T$ (in days), it can be switched to the completely off state. In other words, the completely off state happens when: 
\begin{equation}
	\sum_{t^j_m=1}^{T} x^j_{m,3} D^j_m (t^j_m)=0.
\end{equation}
We denote $a^{0,j}_{m,2}$ as an action of switching from deep sleep to the completely off state. Otherwise, this action $\tilde{a}^{0,j}_{m,3}$ is taken, where the microwave radio remains in the deep sleep state. 

Based on the above-defined actions, we define $\mathcal{A}^j=\{a^{b,j}_{m, {k+1}},  a^{b,j}_{m}\}$ as an action space for $b=0, 1$ for physical states of microwave radio and $ k$ for states of microwave radios at the controller. Furthermore, we define $\Phi(\mathcal{S}^j, \mathcal{A}^j)$ as a transition matrix from one state to another, which depends on states $\mathcal{S}^j$ and actions $\mathcal{A}^j$.

To reduce computational overload on controllers at each microwave node and IAB donor, we propose DT, which represents the microwave backhaul connecting the IAB-based FWA network to the CN. The DT is constructed using microwave backhaul data and deployed on the MEC server in the CN. After DT construction, state transitions $\mathcal{S}^j$ and actions $\mathcal{A}^j$ are computed within the DT and then transmitted to the microwave node and IAB donor for acting. As illustrated in Fig.~\ref{SystemModel2}, the DT in the MEC server consists of two main modules: the Data Acquisition module and the Think, Act, and Plan module, each with several sub-modules:
\begin{itemize}
	\item Network Representation, Service Level Agreement (SLA), and Key Performance Indicator (KPI) Definition Sub-module:  
	This sub-module collects information on microwave nodes and IAB donor, such as available microwave radios, and their configurations (e.g., frequency bands). This process is used to create or update the DT model of the microwave backhaul.   
	\item Nodes, Links, and Network Demand Monitoring Sub-module: 
	This sub-module gathers performance-related data, including data rate, delay, transmission power, energy consumption, microwave radio utilization, and radio states. This process, referred to as network sensing, ensures that the collected information is then passed to the Think, Act, and Plan module.   
	\item KPI and SLA Evaluation Sub-module:
	The monitored data are compared against predefined KPIs and SLAs (e.g., throughput requirements, latency constraints, and power consumption targets). If the performance meets the KPIs and SLAs, the DT instructs the Physical Twin (PT) to maintain the current microwave radio states.   
	\item Parameter and Self-Healing Sub-module: 
	If network performance does not meet the KPIs and SLAs, the DT adjusts the microwave radio states accordingly to satisfy them. Once adjustments are applied, the DT validates whether the new states satisfy the KPIs and SLAs. If successful, the DT sends the adjusted states to the PT for implementation.   
	\item Maintenance Plan Sub-module:
	If the adjusted states still fail to meet the KPIs and SLAs, the DT activates the maintenance plan, which may involve scheduling a site visit for a technician to inspect the microwave backhaul and make proper setup and configuration to meet the KPIs and SLAs.  
\end{itemize}

\section{Problem Formulation}
 \label{sec:problem_formulation}
\begin{figure}[t]
	\centering
	\includegraphics[width=1.0\columnwidth]{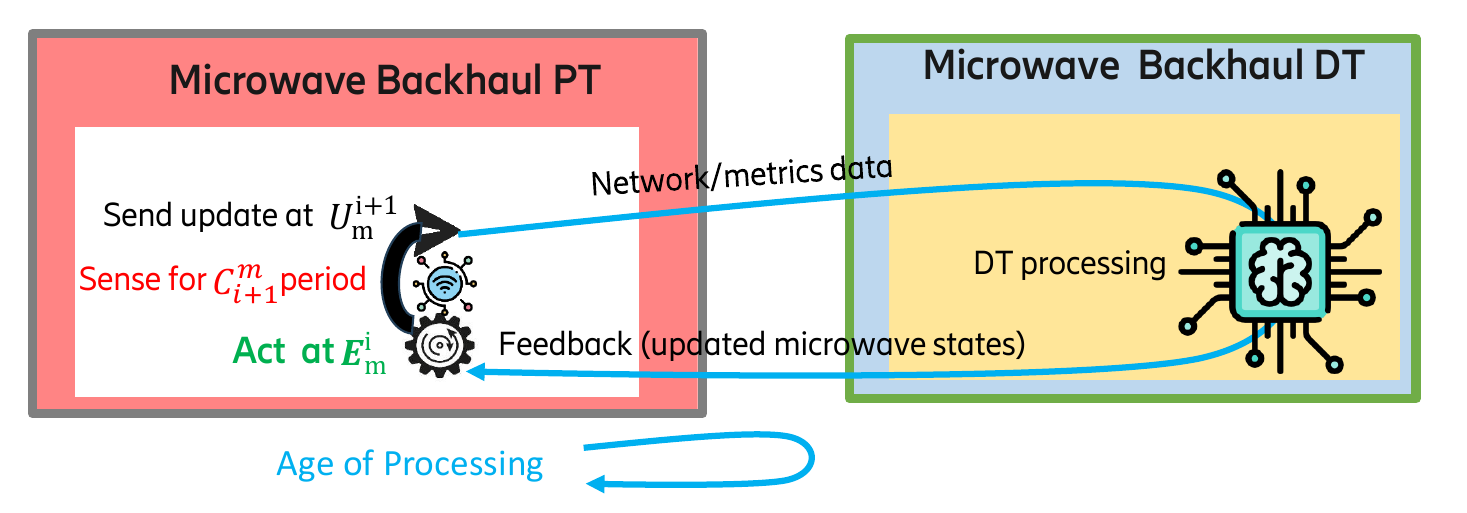}
	\caption{AoP and interaction between PT and DT.}
	\label{PT_DT_Interaction2}
\end{figure}

\begin{figure}[t]
	\centering
	\includegraphics[width=1.0\columnwidth]{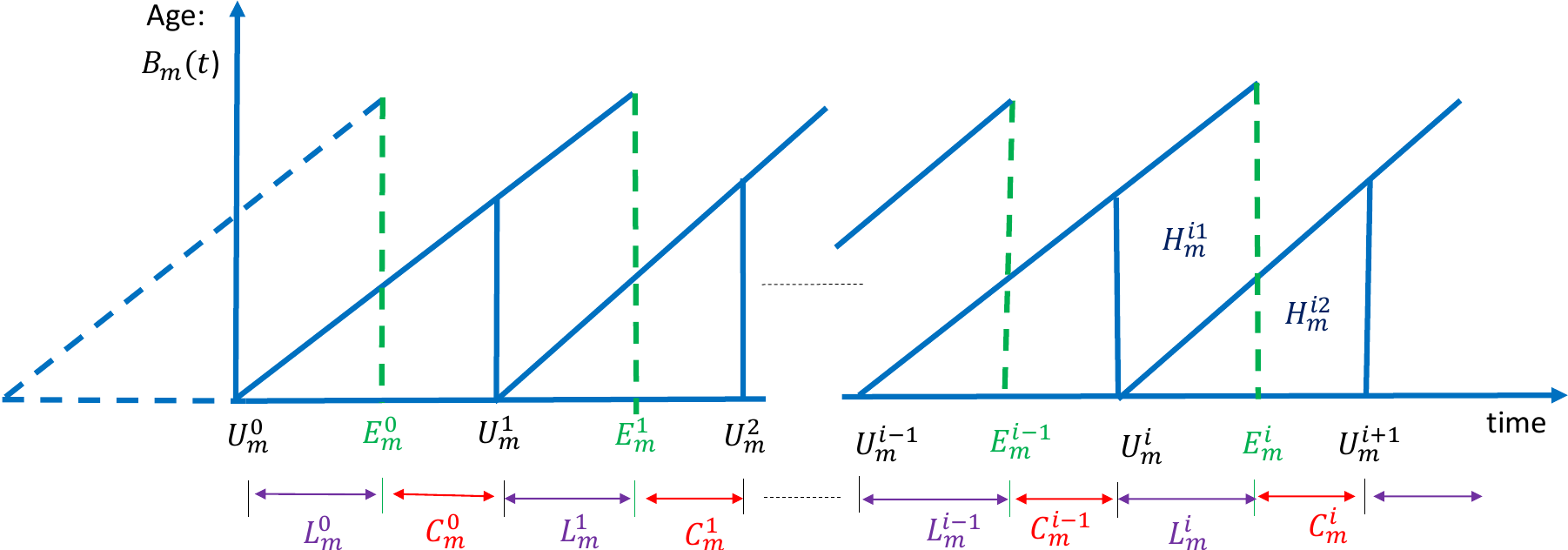}
	\caption{AoP modeling for PT.}
	\label{AoP_offlaoding}
\end{figure}

To reduce both computation and network delay in the microwave backhaul between DT and PT, we consider AoP. In other words, AoP combine computation delay with  the total network delay (i.e.,  processing delay, queuing delay, transmission delay, and propagation delay). As shown in Figs. \ref{PT_DT_Interaction2} and \ref{AoP_offlaoding}, at PT, we define the total time for sending update $i$ (collected network information) to DT and receiving feedback (optimized states and actions) from DT as $E_m^i$. Also, we define $L_m^i$ as the period between sending update $i$ to DT and receiving DT's feedback on the already submitted update at time $U_m^{i-1}$. By considering computation at DT and network delay, the feedback on submitted update $i$ is received at the microwave node or the IAB donor that uses microwave radio $m$ at time:
\begin{equation}
	\begin{aligned}
		& E_m^i = U_m^i +L_m^i.
	\end{aligned}
	\label{eq:get_results}
\end{equation} 
Furthermore, we assume that the PT continues to collect backhaul network metrics. Therefore, from $E_m^i$, the PT generates a new update to send to DT after time $C_m^i \geq 0$. The PT samples new status update $i+1$  at time $U_m^{i+1}$, where $U_m^{i+1}$  is given by:
\begin{equation}
	U_m^{i+1}= E_m^i+C_m^i.
\end{equation}


Time of the freshest status update $i$ at the PT node using microwave radio $m$  is given by:
\begin{align}
	U_m=\max \{U_m^i, E_m^i \leq t; \forall i\}
	\label{eq:age_of},
\end{align}
where time $t$ (with $t \neq t^j_m$) is the current time. From (\ref{eq:age_of}), we can calculate instantaneous AoP $B_m(t)$ of new status update $i+1$ at time $t$ as follows:
\begin{equation}
	B_m(t) = t-U_m.
\end{equation}
In other words, $B_m(t)$ is expressed as a time period. Then, we can use $B_m(t)$ to compute the AoP  $B_m$, where $B_m$ is expressed as:
\begin{equation}
	\label{eq:proce_time}
	B_m=\lim_{t\to\infty} \frac{1}{t} \int_{0}^{t} B_m(t)dt.
\end{equation}
We simplify (\ref{eq:proce_time}) by decomposing the integral into a series of areas of parallelograms and triangles as shown in Fig. \ref{AoP_offlaoding}. The area of parallelogram $H_m^{i1}$ can be expressed as follows:
\begin{equation}
	\label{eq:Q1}
	H_m^{i1}=(L_m^{i-1} + C_m^{i-1})L_m^i,
\end{equation}
while the area of triangle $H_m^{i2}$ can be defined as follows:
\begin{equation}
	\label{eq:Q2}
	H_m^{i2}=\frac{1}{2}(L_m^{i} + C_m^{i})^2.
\end{equation}
Therefore, by considering (\ref{eq:Q1}) and (\ref{eq:Q2}), the average AoP $\tilde{B}_{m}$ is given by:
\begin{equation}
	\label{eq:averageAoP1}
	\tilde{B}_{m}=\frac{\sum_{m\rightarrow \infty}H_m^{i1}+H_m^{i2}}{\sum_{m\rightarrow \infty}L_m^{i} + C_m^{i}}.
\end{equation}

Based on the average AoP $\tilde{B}_{m}$, we formulate the following optimization problem that aims to minimize energy consumption while satisfying the IAB-based FWA’s performance constraints. Energy consumption is modeled as the product of the power consumption $P^{j}_{m,k}$, which depends on the microwave states, and the AoP. Since  power consumption is measured in watts and AoP in seconds, the resulting energy is expressed in joules:
\begin{subequations}
	\label{eq:problem_formulation4}
	\begin{align}
		&\underset{\vect{x}}{\text{min}}
	\sum_{k=0}^{K}\sum_{m=1}^{M^j}
	 x^j_{m,k}
	\left\|
	\Phi(\mathcal{S}^j,\mathcal{A}^j)
	\right\|_{F}
	P^{j}_{m,k}\tilde{B}_{m}
		\tag{\ref{eq:problem_formulation4}}\\
		& \text{subject to: }\nonumber\\
		&\sum_{k=6}^{K}\sum_{m=1}^{M^j}x^j_{m, k} P^{TX}_{m,k} \leq\sum_{k=6}^{K}\sum_{m=1}^{M^j}\tilde{P}^{TX}_{m,k},\label{first:a1}\\
		&\sum_{k=6}^{K}\sum_{m=1}^{M^j}D^j_m\geq D,\label{first:a2}\\
		&\sum_{k=6}^{K}\sum_{m=1}^{M^j}	 x^j_{m, k} \geq 1, \label{first:a3}\\ 
		& x^j_{m, k} \in \{0,1\}. \label{first:a4}
	\end{align}
\end{subequations}

The objective function above applies to each microwave node and IAB donor and depends on the number of states and microwave radios, since each microwave radio does not operate independently.
Constraint (\ref{first:a1}) ensures that the transmission power does not exceed the maximum allowable power limit $\tilde{P}^{TX}_{m,k}$ for microwave radios in serving state. Constraint (\ref{first:a2}) guarantees that the microwave radios in serving state should support the required downlink data rate for IAB-based FWA. Constraint (\ref{first:a3}) enforces that each microwave node and IAB donor has at least one radio in the serving state. Finally, Constraint (\ref{first:a4}) ensures that $x^j_{m, k}$ is binary decision variable.
 \begin{figure*}[t]
	\centering
	\includegraphics[width=2.0\columnwidth]{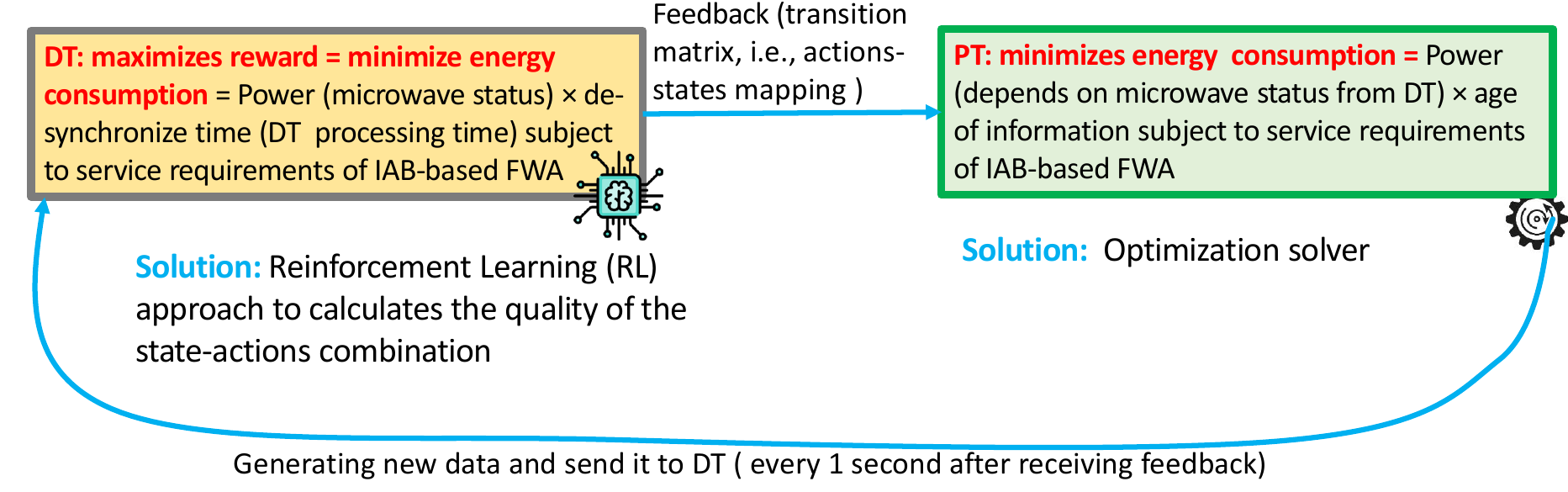}
	\caption{Summary of the proposed solution approach for (\ref{eq:de-synchronization112}) and (\ref{eq:problem_formulation4}).}
	\label{PT_DT_Interaction}
\end{figure*}
		 
\section{Solution Approach}	
\label{SolutionApproach}

In (\ref{eq:problem_formulation4}), the transition matrix $\Phi(\mathcal{S}^j, \mathcal{A}^j)$ can, in principle, be obtained from historical state–action data of microwave radios. However, such data is not available in practice, making the transition matrix unknown. Moreover, estimating $\Phi(\mathcal{S}^j, \mathcal{A}^j)$ locally at each microwave node or IAB donor would incur significant computational overhead.
	To address this challenge, we leverage DT deployed at the MEC server in the CN to estimate the transition matrix. Specifically, the DT continuously collects real-time data from the PT and constructs a virtual representation of the network. Based on this virtual model, the DT learns and updates microwave state transition behaviors, i.e., $\Phi(\mathcal{S}^j, \mathcal{A}^j)$, and enables safe evaluation of state-control decisions before their deployment in the PT. Placing the DT at the MEC server further reduces the computational burden on individual microwave nodes and IAB donor, enabling efficient, adaptive, and SLA-compliant decision-making in a centralized manner.
	
	The PT continuously monitors the network, collects runtime measurements, and transmits them to the DT. The DT processes this information and evaluates $\Phi(\mathcal{S}^j, \mathcal{A}^j)$, which is then used to guide control decisions in the PT. The data acquisition framework includes three key components: (i) the backhaul network representation, which defines the virtual model of the PT, (ii) the service level agreement (SLA), which specifies the required data and service constraints for CPEs/MTs, and (iii) the key performance indicators (KPIs), which define measurable performance targets.
	
	Within the DT, the collected runtime metrics from the PT include radio states, achieved data rates, availability, power consumption, throughput, and resource utilization. These measurements are compared against predefined SLA and KPI requirements. Accordingly, we define a decision variable $y_m$ for each microwave radio $m$ as:
	\begin{equation}
		\setlength{\jot}{10pt}
		y_m =
		\begin{cases}
			1, \text{if PT monitoring results satisfy KPI and SLA,} \\
			0, \text{otherwise.}
		\end{cases}
	\end{equation}
	
	If $y_m = 1$, the DT confirms that the current configuration of microwave radio is satisfactory and instructs the PT to maintain the existing microwave radio states while continuing data collection. Otherwise ($y_m = 0$), the DT explores alternative configurations, such as putting in serving state a microwave radio from deep sleep or turning on a completely off radio, in order to satisfy SLA and KPI requirements. These updated configurations are used to recompute $\Phi(\mathcal{S}^j, \mathcal{A}^j)$ within the DT environment before applying it to PT. The configurations are validated within the DT, leading to a second decision variable $z_m$ defined as:
	\begin{equation}
		\setlength{\jot}{10pt}
		z_m =
		\begin{cases}
			1, & \text{if configuration at the DT using } \Phi(\mathcal{S}^j, \mathcal{A}^j) \\
			& \text{ satisfies KPI and SLA requirements,} \\
			0, & \text{otherwise.}
		\end{cases}
	\end{equation}

During the computation and testing of $\Phi(\mathcal{S}^j, \mathcal{A}^j)$ at the DT, a synchronization gap may arise between PT and the DT due to processing delays at the MEC server. We define the resulting desynchronization time $\Delta_\tau$ as:
	\begin{equation}
		\label{eq:de-synchronization}
		\Delta_\tau=\frac{\tilde{D}\Lambda_{mec}}{f_{mec}},
	\end{equation}
where $f_{mec}$ denotes the CPU frequency of the MEC server, $\Lambda_{mec}$ represents the number of CPU cycles required for DT processing, and $\tilde{D}$ is the size of the collected network/monitoring data. Furthermore, the energy consumption associated with DT processing at the MEC server is given by:
	\begin{equation}
		L_{mec}= \tilde{D} \, \iota_{mec}\Lambda_{mec} f_{mec}^2,
	\end{equation}
where $\iota_{mec}$ is a constant determined by the MEC CPU architecture. Based on these definitions, we define  reward function $R(\mathcal{S}^j, \mathcal{A}^j)$ in DT as:
	\begin{equation}
		\label{eq:de-synchronization111}
		R(\mathcal{S}^j, \mathcal{A}^j)=
		\frac{\sum_{m=1}^{M^j} \left(y_m+ z_m\right)}
		{\sum_{k=0}^{K}\sum_{m=1}^{M^j}
			\left\|
			\Phi(\mathcal{S}^j,\mathcal{A}^j)
			\right\|_{F}
			P^{j}_{m,k} \Delta_\tau
			+ L_{mec}}.
	\end{equation}
	
Maximizing $R(\mathcal{S}^j, \mathcal{A}^j)$ also helps to minimize the total energy cost in the denominator. Specifically, the denominator captures both (i) the communication-related energy consumption, which depends on the power consumption determined by microwave radio states and desynchronization time $\Delta_\tau$, and (ii) the computational energy consumption $L_{mec}$ at the MEC server. Moreover, even when no CPE or IAB-MT is active, microwave nodes and IAB donors still consume a non-zero amount of power due to control signaling operations. Therefore, the denominator of (\ref{eq:de-synchronization111}) never becomes zero, ensuring that the reward function remains well-defined. Finally, if $R(\mathcal{S}^j, \mathcal{A}^j)=0$, then $y_m + z_m = 0$, meaning that neither the PT nor the DT can satisfy the SLA and KPI requirements. Therefore, a maintenance plan should be considered, which may require physical intervention.

At DT, maximizing $R(\mathcal{S}^j, \mathcal{A}^j)$ requires to estimate $\Phi(\mathcal{S}^j, \mathcal{A}^j)$, where we use a Reinforcement Learning (RL) approach \cite{ernst2024introduction} to approximate  $\Phi(\mathcal{S}^j, \mathcal{A}^j)$. In particular, we use Deep Q-Learning (DQL)  \cite{clifton2020q}, which evaluates the quality of state-action pairs through the Q-function:
\begin{equation}
		\label{eq:de-synchronization112}
	Q:\mathcal{S}^j \times \mathcal{A}^j \rightarrow \mathcal{R}(\mathcal{S}^j, \mathcal{A}^j).
\end{equation}  
We chose DQL over other approaches because the transition matrix for the microwave node and IAB donor is unknown, making classical optimization intractable. As a model-free method, DQL learns optimal state–action policies directly from interactions between the DT and PT without requiring an explicit transition model. 

As illustrated in Fig.~\ref{PT_DT_Interaction}, the transition matrix $\Phi(\mathcal{S}^j, \mathcal{A}^j)$ is obtained from DQL in maximizing $R(\mathcal{S}^j, \mathcal{A}^j)$ and transmitted as feedback to the PT. This feedback enables each microwave node and IAB donor to solve (\ref{eq:problem_formulation4}) with known  $\Phi(\mathcal{S}^j, \mathcal{A}^j)$. In other words, the problem (\ref{eq:problem_formulation4}) can be solved using standard solvers such as Gurobi or CPLEX. After solving (\ref{eq:problem_formulation4}), each microwave node and IAB donor collect new network metrics and transmit them to the DT  as a status update. In the proposed solution, we use $C_m^i = 1$ second; however, this choice is not restrictive, and other update intervals can be applied. Nonetheless, selecting a larger value of $C_m^i$ may prevent the DT from accurately reflecting the PT’s real-time status.

\begin{remark}
	\emph{Computational complexity of the proposed solution approach is $\mathcal{O}(n^2)$.} 
\end{remark}
In the DT, DQL employs a deep neural network (DNN) \cite{mnih2015human, van2016double} to approximate the Q-value function $Q$ in (\ref{eq:de-synchronization112}). The primary computational overhead originates from the DNN training phase, where each parameter update requires both forward and backward propagation through the network. The corresponding computational complexity can be expressed as
$\mathcal{O}\!\left(\mu \sum_{l=1}^{L} n_{l-1}n_l\right)$, $\mu$ denotes the minibatch size, $L$ is the number of neural network layers, and $n_l$ represents the number of neurons in layer $l$. Therefore, the complexity increases with the DNN depth and the size of the state-action space. However, in this work, the DNN is trained offline at the DT hosted on the MEC server in the core network (CN). Consequently, the online DT operation performs inference only using the pre-trained model, thereby significantly reducing real-time computational overhead. Moreover, since rural microwave backhaul networks experience relatively slow variations in traffic demand and microwave radio states compared with dense urban deployments, the DT computational requirements remain manageable.

To evaluate whether the estimated transition matrix $\Phi(\mathcal{S}^j,\mathcal{A}^j)$ satisfies the SLA and KPI requirements, the DT performs verification operations over the microwave radio states and monitoring results. Let $\vect{y}$ and $\vect{z}$ denote vectors associated with microwave radio states and SLA/KPI validation indicators, respectively, each having dimension $n$. In the worst case, the DT must compare all possible state-validation combinations, resulting in a computational complexity of $\mathcal{O}(n^2) $.

In the PT of the microwave backhaul, the computational complexity mainly depends on the number of microwave radios and operational states. Let $n$ denote the total number of microwave radio-state combinations. Solving (\ref{eq:problem_formulation4}) requires iterating through the radio-state decision variable $x^j_{m,k}$, resulting in a linear complexity of $\mathcal{O}(n)$.

The total computational complexity of our proposal can be expressed as $\mathcal{O}\!\left(\mu \sum_{l=1}^{L} n_{l-1}n_l
\right)+\mathcal{O}(n^2)+ \mathcal{O}(n)$. Since the DNN training is performed offline, the quadratic verification process dominates the online execution. Therefore, the overall worst-case computational complexity of the proposed solution can be approximated as $\mathcal{O}(n^2)$.
\begin{figure}[t]
	\centering
	\begin{minipage}{0.45\textwidth}
		\centering
		\includegraphics[width=1.0\columnwidth]{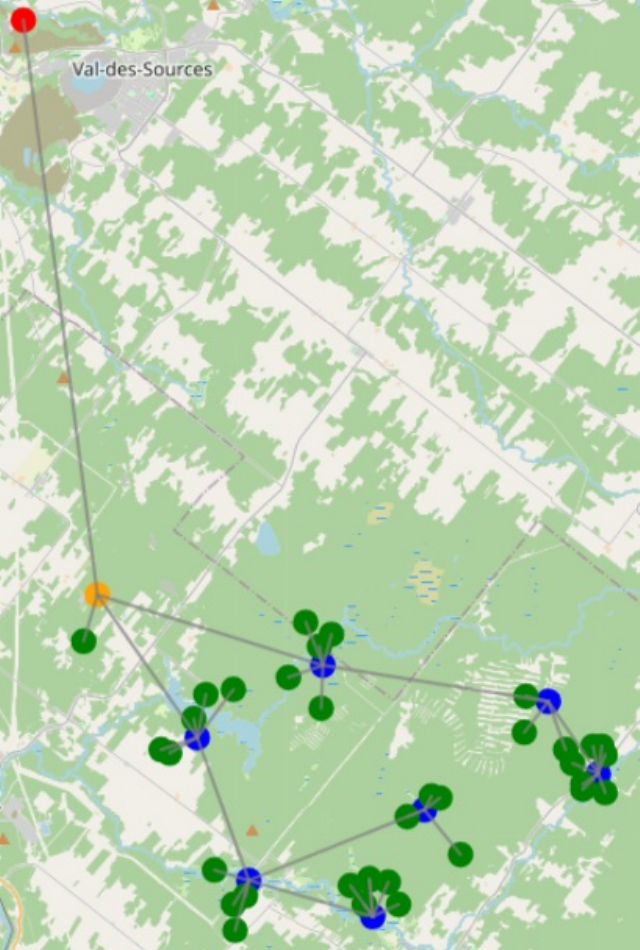}
	\caption{ Nodes in rural area under consideration.}
	\label{fig:TopologyMap}
\end{minipage}
\end{figure}
\begin{figure}[t]
	\centering
	\begin{minipage}{0.45\textwidth}
		\centering
		\includegraphics[width=1.0\columnwidth]{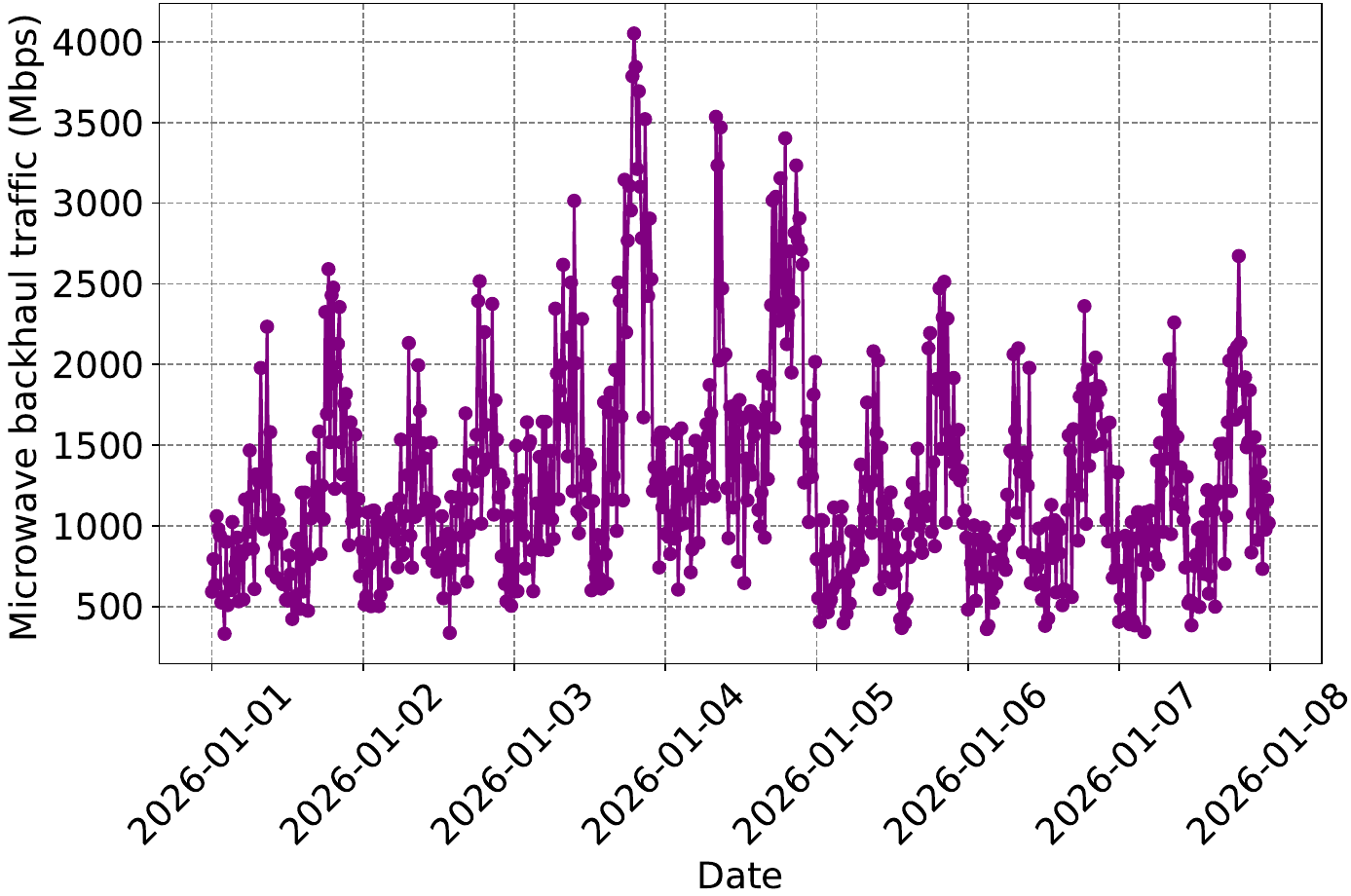}
		\caption{ Downlink microwave backhaul data for IAB-based FWA.} 
		\label{fig:BackhaulTraffic}
	\end{minipage}
\end{figure}

\section{Simulation Results and Analysis}
\label{sec:PerformanceEvaluation}
\begin{figure}[t]
	\centering
	\begin{minipage}{0.45\textwidth}
		\centering
		\includegraphics[width=1.0\columnwidth]{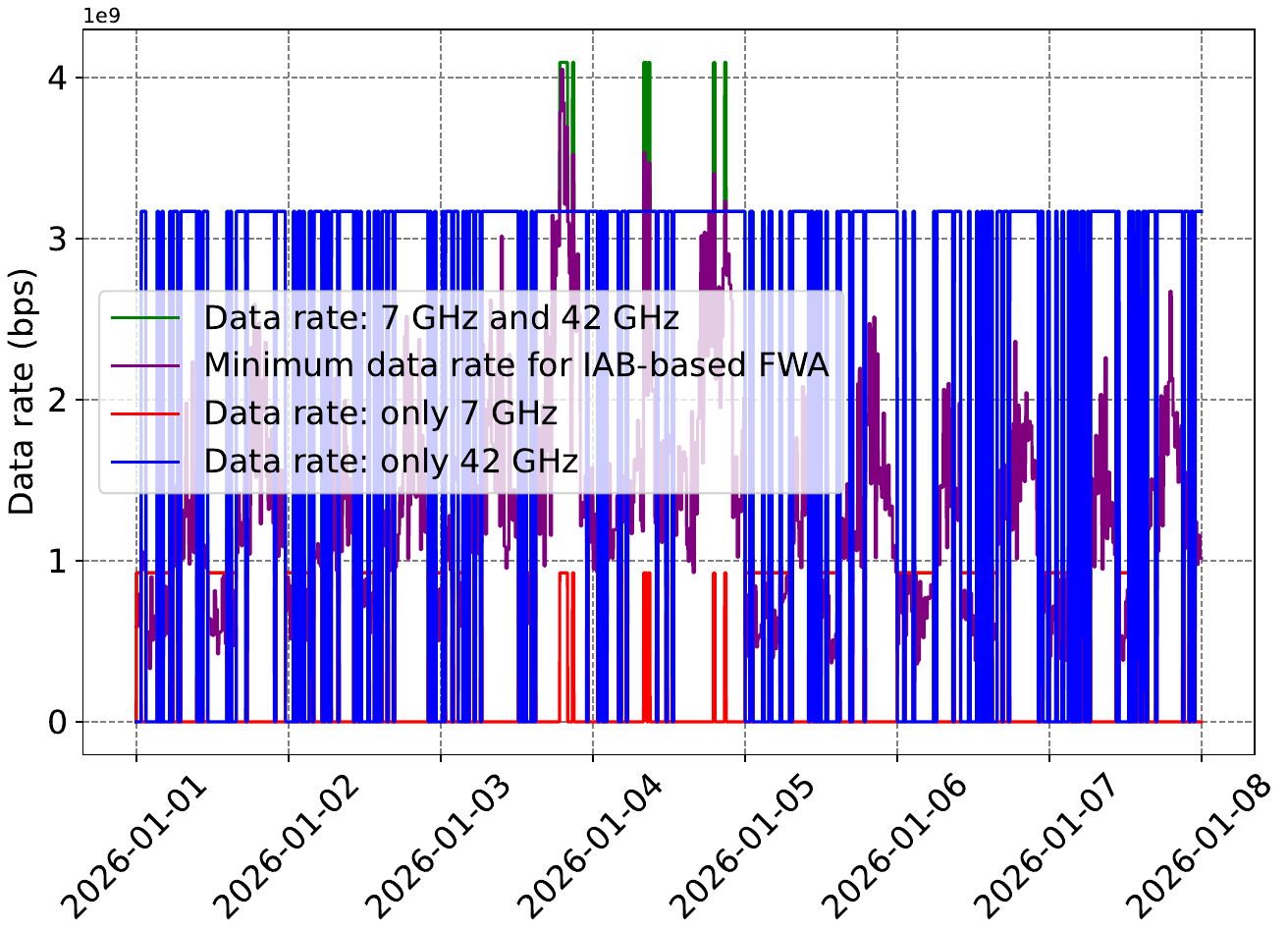}
		\caption{Data rate requirements with radio states.} 
		\label{fig:DataRateState}
	\end{minipage}
	\begin{minipage}{0.45\textwidth}
		\centering
		\includegraphics[width=1.0\columnwidth]{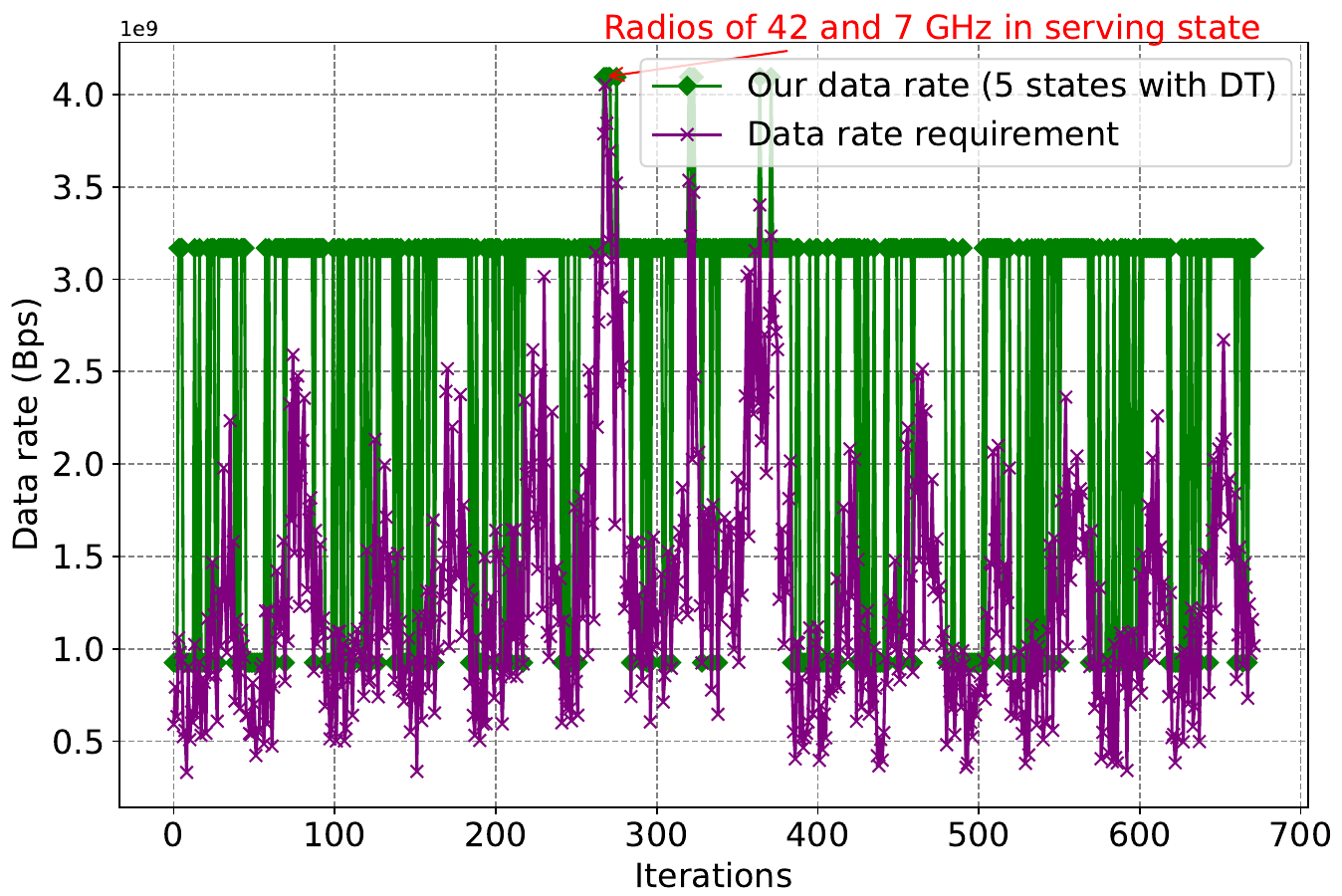}
		\caption{Our approach in satisfying data rate requirements.} 
		\label{fig:DataRate}
	\end{minipage}
\end{figure}

In this section, we describe the simulation setup and present the results of the performance evaluation. The simulation environment is implemented using Python $3.8.10$ for numerical analysis and CVXPY \cite{diamond2016cvxpy} with embedded conic solver (ECOS) for solving optimization problem in PT. Furthermore, for DQL in DT, we use PyTorch \cite{ketkar2021introduction}.

\subsection{Simulation Setup}
\label{subsec:SumulationSetup}

To construct the network topology, we randomly selected a rural area in Quebec, Canada, as illustrated in Fig. \ref{fig:TopologyMap}. The microwave node (i.e., the red node) is located near the town of Val-des-Sources. It is connected to an IAB donor (node in yellow) located $17.95$ km away, providing service coverage for $35$ houses (i.e., CPEs are the nodes in green) in the surrounding rural area.

For the microwave node and the IAB donor, each is equipped with two radios: one operating in the 7~GHz band with 64~MHz bandwidth, and another in the 42~GHz band with 500~MHz bandwidth. The power states of the microwave radios are modeled as follows: $0$~W (completely off), $3$~W (deep sleep), $55$~W (startup), $50$~W (wake-up), and $80$~W (serving). In practice, however, these power levels can be directly measured from the microwave equipment. 
Furthermore, we set $t^j_{m,4} = 10$~seconds and $t^j_{m,5} = 8$~seconds. These values are reasonable for microwave backhaul radios, which typically maintain stable operation by aggregating traffic from multiple access radios, thereby enabling continuous traffic availability. As a result, frequent state transitions at a second or millisecond scale during the day are uncommon, an assumption supported by~\cite{frithiofson2022energy, ericssonpower}. However, these parameter values are not suitable for IAB-based FWA radios that operate under different resource allocations.

In IAB-based FWA, 7 IAB nodes (i.e., nodes in blue in Fig. \ref{fig:TopologyMap}) are connected to the IAB donor to serve 35 houses with CPEs. As the signal attenuates with distance, we assume each CPE is connected to the nearby IAB station. Nearby CPEs or IAB-MTs to the parent IAB station operated in the mmWave band, while distant ones used the mid-band.   The IAB-based FWA network operated at $38$ $GHz$ for mmWave and $6$ $GHz$ for mid-band, with bandwidths within the range of $10$ to $1600$ $MHz$ and subcarrier spacing in the range $15$ to $480$ kHz \cite{etsi138}. The results of the use of mmWave and mid-band for IAB-based FWA  are discussed in our previous work in \cite{ndikumana2025energy, ndikumana2026energy}. In this performance evaluation, we focus on the use of microwave backhaul to support IAB-based FWA.

The network traffic in the considered rural scenario is synthetically generated over a one-week period with a temporal resolution of one second. The generated traffic model captures both short-term variations and recurring daily patterns, providing a realistic and flexible representation of aggregated data demand for performance evaluation in rural environments. As expected in rural access networks, traffic demand is higher during daytime periods and significantly lower during nighttime hours. The generated traffic profile is then used as an input to the proposed approach, which is designed to support arbitrary and dynamic traffic patterns. The resulting backhaul traffic demand for the IAB-based FWA scenario is presented in Fig.~\ref{fig:BackhaulTraffic}.

\subsection{Simulation Results}

Considering the five operational states of each microwave radio, where each microwave node and IAB donor are equipped with two microwave radios ($M^j = 2$) like the ones shown in Fig. \ref{fig:microwavetwol}. The DT  decides on states to minimize energy consumption while satisfying the required microwave backhaul data rate. As illustrated in Figs \ref{fig:DataRateState} and \ref{fig:DataRate}, when the data rate requirement is relatively low, a  $42$ $GHz$ microwave radio with $500$ $MHz$ bandwidth is sufficient to meet the backhaul data rate requirement. In this case, the $7$ $GHz$ microwave radio with $64$ $MHz$ bandwidth can enter a deep-sleep state to reduce energy consumption. Conversely, when the demand exceeds the capacity of the 42 GHz radio, the 7 GHz radio must wake up to support the 42 GHz microwave radio and ensure that the data rate requirements are met. When both the $42$ $GHz$ and $7$ $GHz$ microwave radios are in serving states, a rate of $4.2$ $Gbps$ can be achieved for microwave backhaul.
\begin{figure}[t]
	\centering
	\begin{minipage}{0.45\textwidth}
		\centering
		\includegraphics[width=1.0\columnwidth]{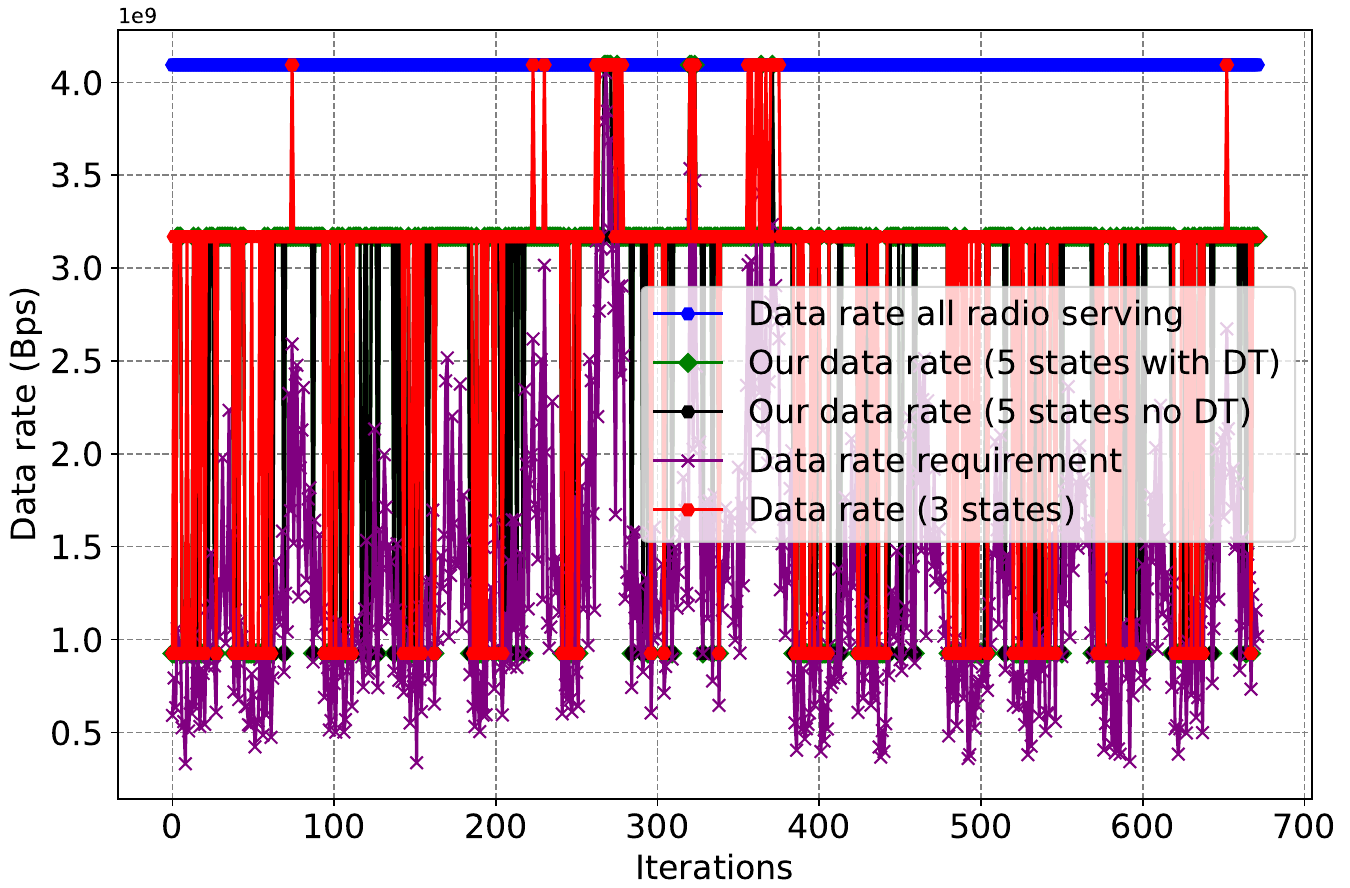}
		\caption{Comparison of our approach with other methods.} 
		\label{fig:DataRateComparison}
	\end{minipage}
	\begin{minipage}{0.45\textwidth}
		\centering
		\includegraphics[width=1.0\columnwidth]{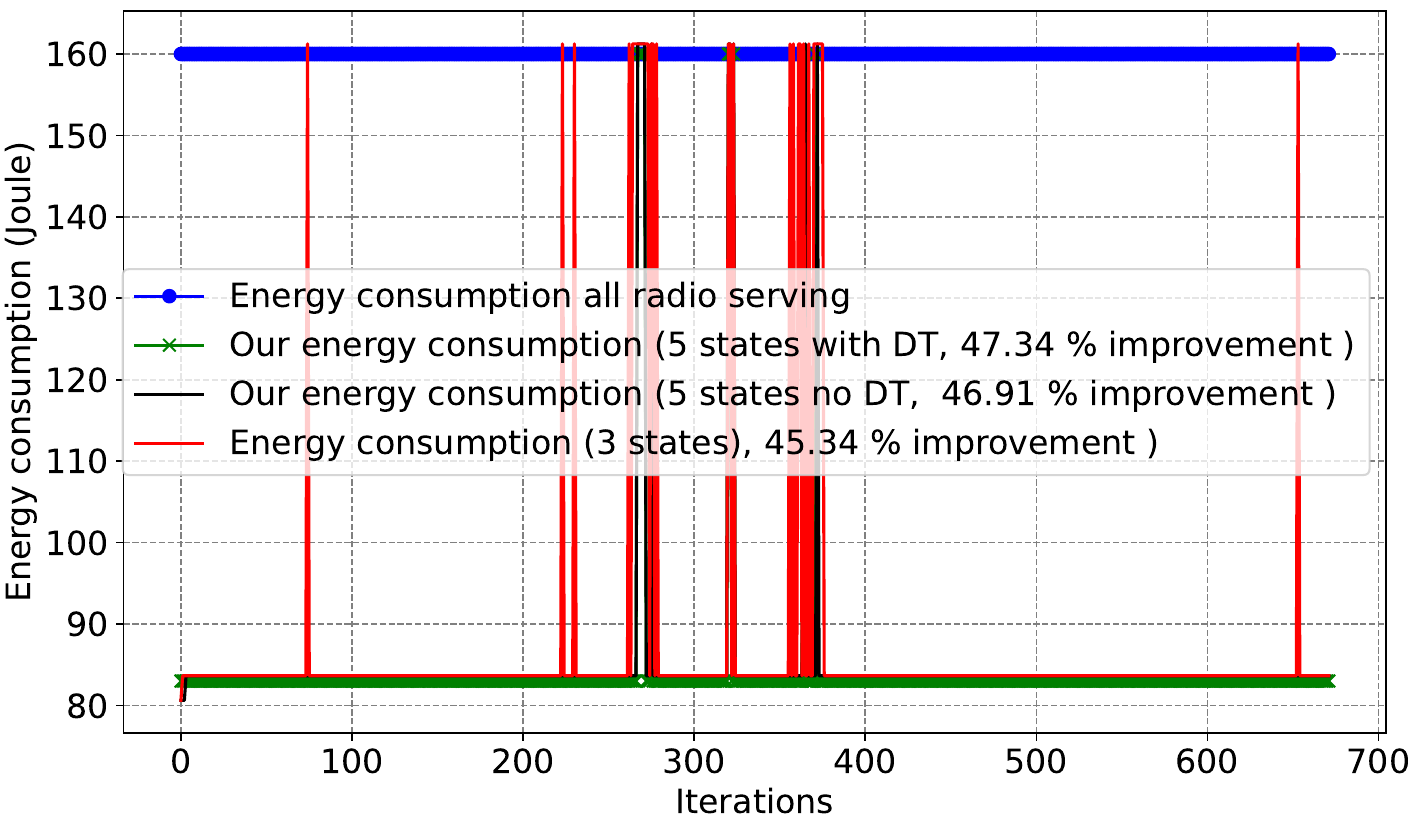}
		\caption{Comparison of energy consumption.} 
		\label{fig:EnergyConsumption}
	\end{minipage}
\end{figure}
\begin{figure}[t]
	\centering
	\begin{minipage}{0.45\textwidth}
		\centering
		\includegraphics[width=1.0\columnwidth]{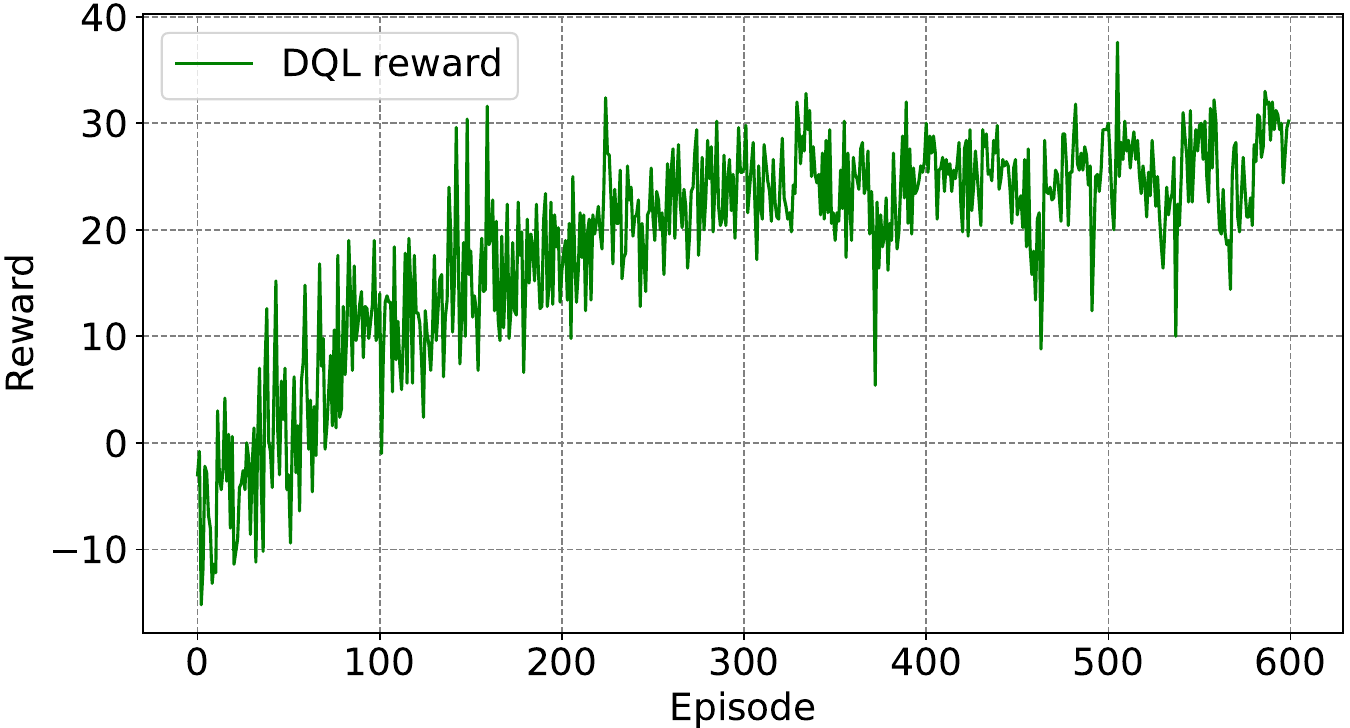}
		\caption{Reward maximization at DT.} 
		\label{fig:Rewards}
	\end{minipage}
	\begin{minipage}{0.45\textwidth}
		\centering
		\includegraphics[width=1.0\columnwidth]{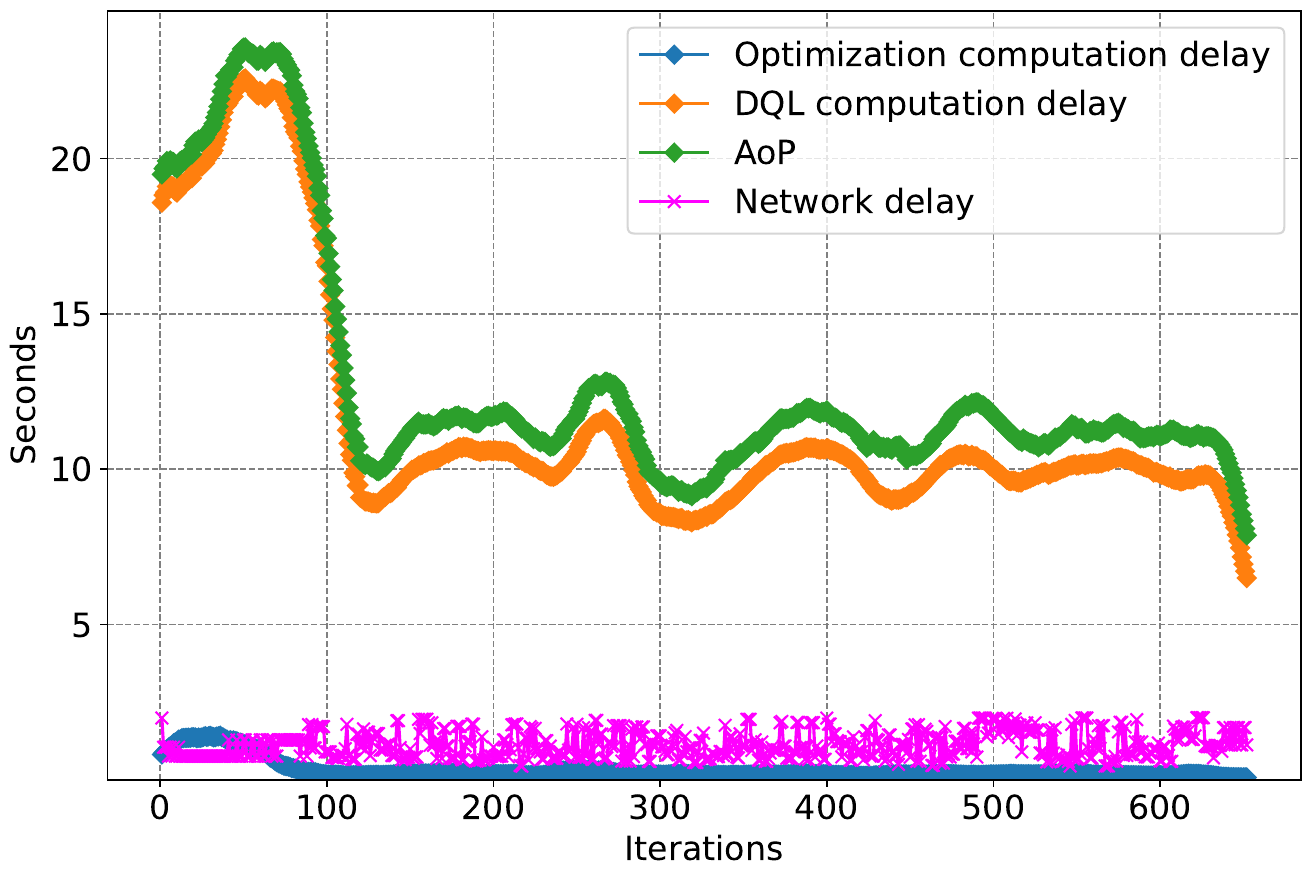}
		\caption{Age of processing and delay.} 
		\label{fig:AoP}
	\end{minipage}
\end{figure}
\begin{figure}[t]
	\centering
	\begin{minipage}{0.45\textwidth}
		\centering
		\includegraphics[width=1.0\columnwidth]{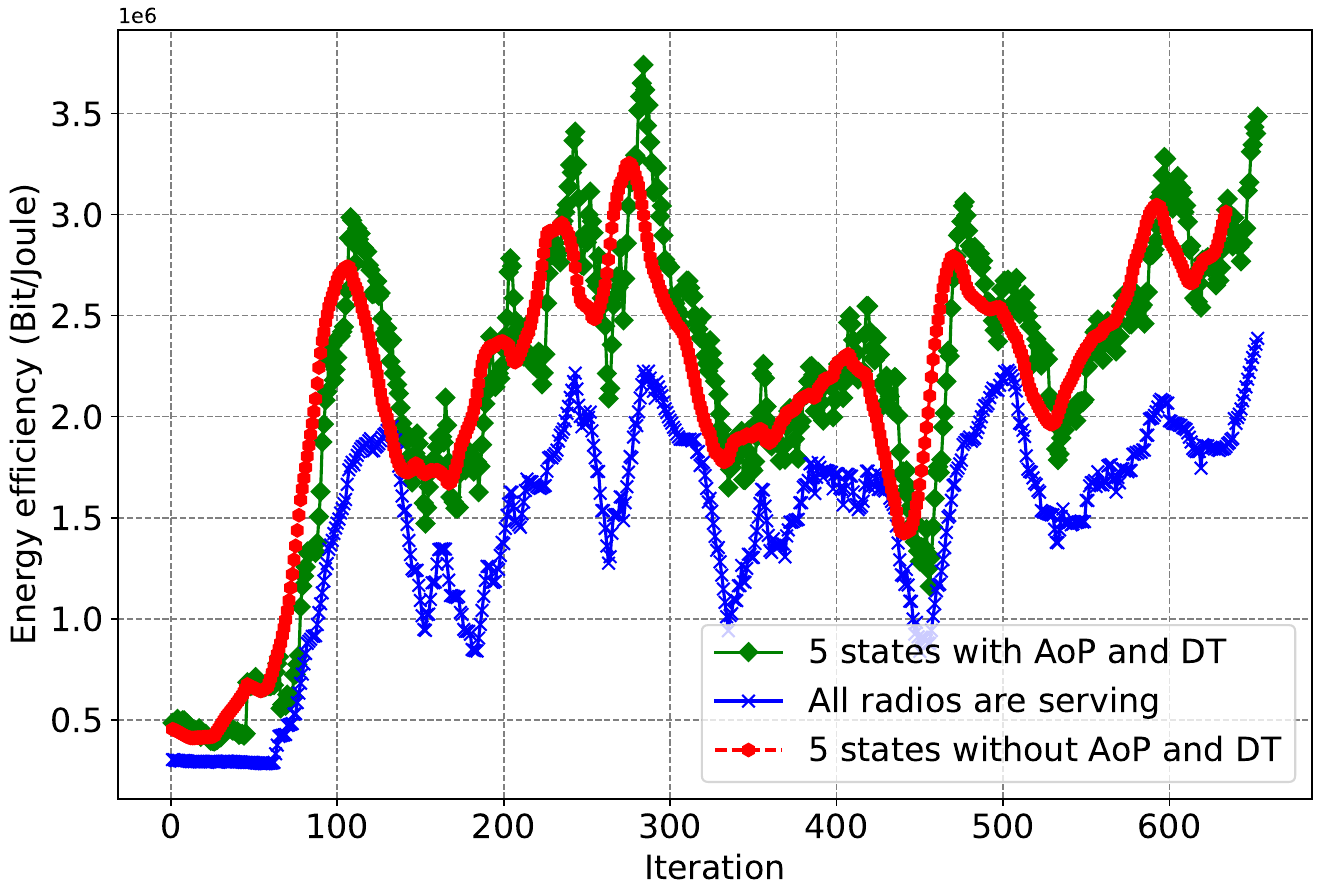}
		\caption{Comparison of energy efficiency.} 
		\label{fig:EnergyEfficiency}
	\end{minipage}
	\begin{minipage}{0.45\textwidth}
		\centering
		\includegraphics[width=1.0\columnwidth]{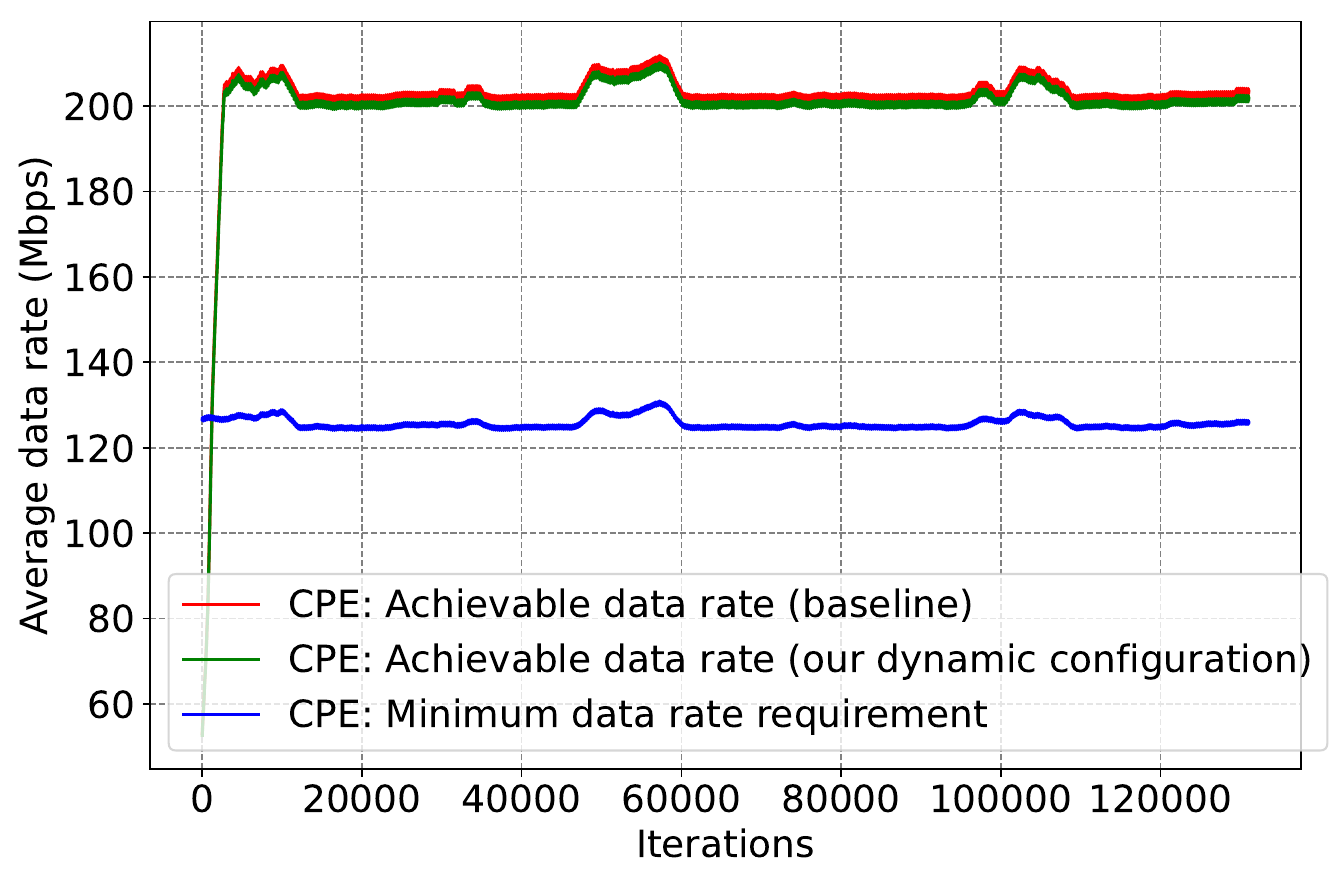}
		\caption{Average achievable data rate per CPE.} 
		\label{fig:DataRateIAB}
	\end{minipage}
	\centering
	\begin{minipage}{0.45\textwidth}
		\centering
		\includegraphics[width=1.0\columnwidth]{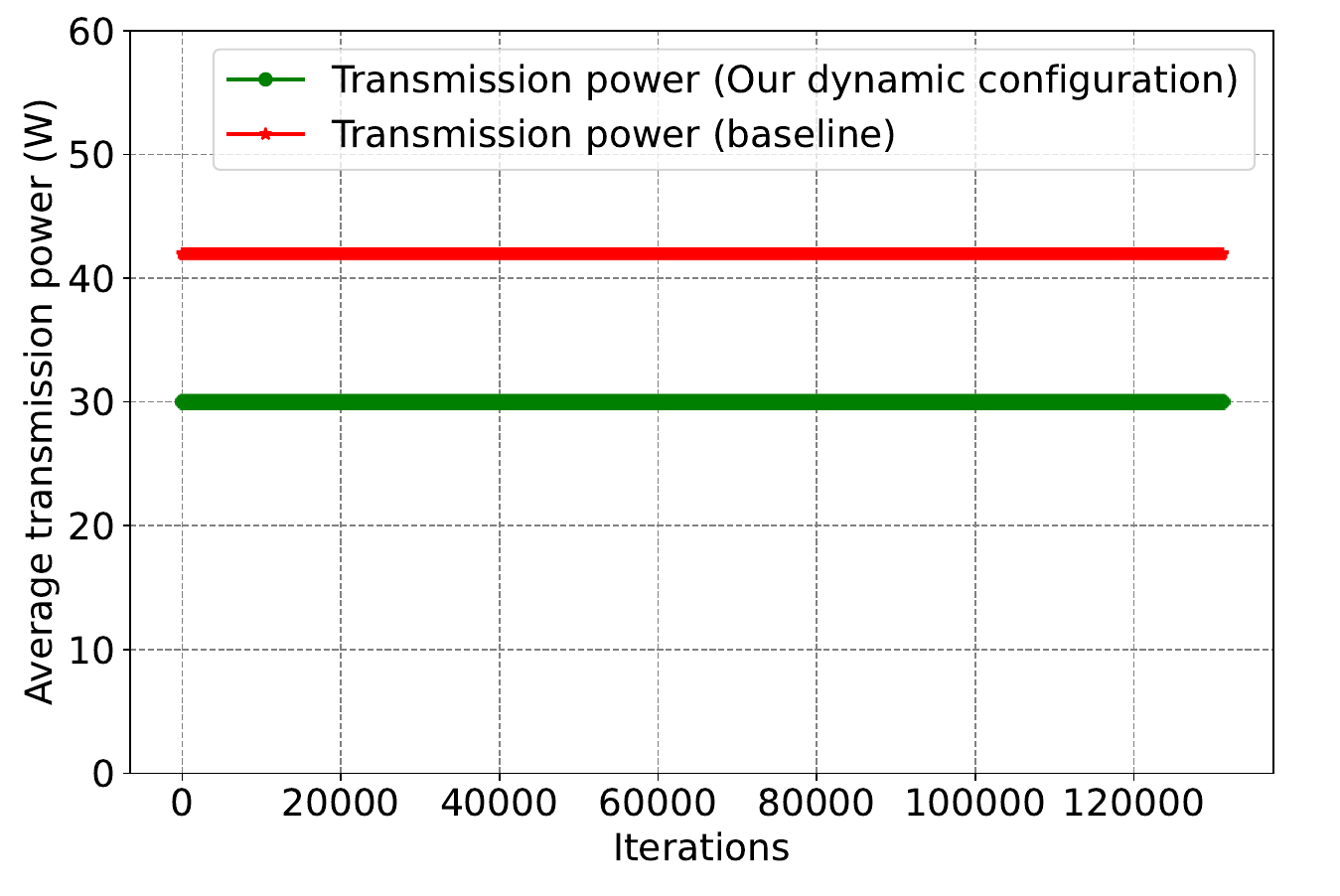}
		\caption{Average transmission power for IAB station.} 
		\label{fig:Transmission_power}
	\end{minipage}
\end{figure}	

In Fig. \ref{fig:DataRateComparison}, we compare our five-state approach with a baseline in  \cite{frithiofson2022energy}. Furthermore, we evaluate the performance of our approach with and without DT support \cite{ndikumana2026energy}. The results demonstrate that our proposal outperforms the three-state baseline in satisfying the data rate requirement of IAB-based FWA. However, when considering our approach with and without DT support, both satisfy minimum data rate requirements.

Fig. \ref{fig:EnergyConsumption} compares the energy consumption of the proposed approach with and without DT support. The results are further benchmarked against a three-state power model and a baseline scenario in which all radios remain continuously serving. The simulation results show that the proposed approach achieves lower energy consumption compared to the baselines. In this figure, energy is expressed in joules; note that $1$~W corresponds to $1$~J/s. The results show an improvement of approximately  $47.34 \%$ over all radios that remain continuously serving and  $2\%$ over the 3-state baseline. This gain can be even more significant when scaled to a network with multiple microwave links, resulting in substantial cumulative energy savings for network operators.

In the DT, we employ DQL to maximize the reward formulated in (\ref{eq:de-synchronization112}). Fig. \ref{fig:Rewards} illustrates the reward maximization process over $700$ episodes. In the context of DQL, an episode is defined as a complete sequence of interactions between the agent and the environment, starting from an initial state and continuing until a terminal state is reached. The DQL model is implemented as a fully connected feedforward neural network. Its architecture comprises two hidden layers with 64 neurons each, with ReLU activation functions to introduce nonlinearity. The final linear layer projects the learned hidden representation onto the action-value space, thereby estimating the Q-values for all possible actions $\mathcal{A}^j$ in a given states $\mathcal{S}^j$. This architecture achieves a balance between simplicity and computation requirement, making it well-suited for reinforcement learning tasks with moderate state and action dimensions.

Considering DQL within the DT and AoP,  Fig.~\ref{fig:AoP} illustrates both the network delay between the PT and DT and the processing time at the DT (DQL computation delay). The figure also shows the computation delay in solving the optimization problem. Notably, distributing DQL to each microwave node and IAB donor without leveraging the DT may consume more energy than executing DQL operations centrally at the MEC server. Considering both data rate and energy consumption, the results in Fig.~\ref{fig:EnergyEfficiency} demonstrate that our proposed approach with DT and AoP achieves higher energy efficiency.

We consider an IAB-based FWA network using 38 GHz for the mmWave band and 6 GHz for the mid-band. A numerology-based resource block allocation scheme from our previous work in \cite{ndikumana2026energy} is employed to select channel bandwidths ranging from $10$ $MHz$ to $1600$ $MHz$ and subcarrier spacings between $15$ $kHz$ and $480$ $kHz$ \cite{etsi138}. The proposed dynamic configuration approach is compared with a baseline scenario that uses fixed numerologies and radio resources, namely 264 resource blocks for the mid-band and 273 resource blocks for the mmWave band, both with a 30 kHz subcarrier spacing \cite{hashemi2017integrated}. Figure \ref{fig:DataRateIAB} illustrates the achievable data rate, while Fig. \ref{fig:Transmission_power} presents the corresponding transmission power. The results show that the proposed dynamic configuration achieves the same data rate as the baseline while requiring lower transmission power.

\section{Conclusion}
\label{sec:Conclusion}
This paper investigated an energy-efficient multi-hop wireless backhaul for rural connectivity by considering long-haul microwave links, 5G IAB, and FWA as a unified solution. To address the overlooked challenge of increasing energy consumption with the number of hops, we modeled the microwave backhaul as a multi-state system, incorporating radio-off, start-up, serving, deep sleep, and wake-up states, and formulated an energy minimization problem that satisfies the data rate requirement of IAB-based FWA serving rural areas. To efficiently solve this problem, we leveraged deep Q-learning within the DT to learn optimal control policies. In contrast, we solve the optimization problem in PT for changing microwave radio states to minimize energy consumption while satisfying data rate requirements. Simulation results demonstrate that the proposed framework achieves substantial energy savings compared to conventional baseline strategies while still satisfying the required data rate constraints for IAB-based FWA. These findings highlight the effectiveness of DT-enabled intelligence for managing energy-performance trade-offs in multi-hop rural wireless networks and provide a promising direction for sustainable, scalable rural broadband deployment.
\bibliographystyle{IEEEtran}

\end{document}